\documentclass[twocolumn,secnumarabic,amssymb, nobibnotes, aps, pra, showpacs,10pt]{revtex4-1}
\pdfoutput=1

\usepackage{amsmath}
\usepackage{amssymb}

\usepackage{graphicx}
\usepackage{natbib}
\usepackage{fancyhdr}
\usepackage[usenames,dvipsnames]{xcolor}
\definecolor{amaranth}{rgb}{0.9, 0.17, 0.31}
\begin{document}

\title{High harmonic generation in an active grating }
\author{C. Chappuis$^{1}$, D. Bresteau$^{1}$, T. Auguste$^{1}$, O. Gobert$^{1}$ and T. Ruchon$^{1,*}$}
\affiliation{$^1$LIDYL, CEA, CNRS, Université Paris-Saclay, CEA Saclay, 91191 Gif-sur-Yvette, France }
\email{thierry.ruchon@cea.fr}

\begin{abstract}
We study theoretically and experimentally High Harmonic Generation (HHG) using two non collinear driving fields focused in gases. We show that these two fields form a non stationary blazed active grating in the generation medium. The intensity and phase structure of this grating rule the far field properties of the emission, such as the relative amplitude of the diffraction orders. Full macroscopic calculations and experiments support this general analysis. This insight into the HHG process allows us to envision new structuration schemes to convert femtosecond lasers to attosecond pulses with increased efficiency. 
\end{abstract}
\maketitle

\section{Introduction}\label{sec:Intro}Attosecond pulses are becoming ultimate tools to address ultrafast processes in the matter, whether in gases or solids. Using \emph{ex situ} pump-probe schemes like transient absorption and photoionization spectroscopies, they are now used to investigate ultrafast photoionization dynamics in atoms and molecules with attosecond resolution  \cite{Calegari2016,Gallmann2017}. As examples of some very recent progress, time delays in photoionization were identified in both atoms and molecules  \cite{SchultzeScience2010,Isinger2017,Vos2018}. Auger decays  \cite{Huetten2018} or the build up of Fano resonances in noble gases were followed in real time  \cite{Kotur2016, Gruson_2016, Cirelli2018}. The deformation of the atomic potential due to the presence of a strong field was measured  \cite{Kiesewetter2017}. Dynamics of holes following photoionization was followed in amino-acids  \cite{CalegariS2014}\ldots In solids, energy-dependent time delays of photoemission, including their variation with the angular momentum of the initial state, or of the final state were measured \cite{Locher2015, Siek2017, Tao2016}. Elastic and inelastic scattering times were identified in dielectric nanoparticles  \cite{Seiffert2017}. Dynamics of magnetization could be probed  \cite{Kfir2017}\ldots As a counterpart of these \emph{ex-situ }schemes, in which attosecond beams are used downstream the generation region to probe matter, HHG also appears as an efficient tool to interrogate highly non-linear processes in matter. Here, HHG radiation from a  pumped sample is analyzed to probe ultrafast processes at play in a so-called \emph{in situ }approach. In particular, the relative phase of concurrent processes triggered by a strong field has been investigated using various schemes, shedding new light on ultrafast non linear processes in atoms and molecules, including chiral molecules and biologically relevant ones  \cite{ItataniNature2004,HaesslerNP2010,CamperPRA2014,FerreNC2015,Baykusheva2017,Kraus2015,Bruner2016,Ferre2016,McGrath2017, Marangos2016, Cireasa2015}. 

Interestingly, for both probing schemes, the use of light induced gratings has lately been promoted. First, the long-known transient grating spectroscopy has been adapted to harmonic spectroscopy, allowing background free detection  \cite{Mairesse2008}. Here, the generating medium is pumped by two beams forming an interference intensity pattern, while a third beam generates harmonics at a delayed time. Second, it was shown that generating harmonics with two beams of equal wavelength crossing in the HHG medium allows imparting a great variety of properties to the outgoing high order harmonics. Here, for a given harmonic number, a series of diffraction orders are observed along the line formed by the difference of the wave vectors of the two beams. This approach, theoretically proposed in Ref. \onlinecite{Birulin1996}, was first experimentally investigated by Bertrand et al. \cite{BertrandPRL2011} and further analyzed by Heyl et al.  \cite{Heyl2014}. It was realized that conservation of parity, energy and linear momentum play a central role to explain the presence/absence of certain channels.  In brief, for harmonic q, its properties are given by the sum of the properties of the n photons absorbed or stimulated emitted from the first beam and m photons of the second beam, where q=m+n. Due to linear momentum conservation, these different channels are spatially separated in the far field and readily identifiable, making attractive sources of tailored attosecond pulses. These conservation rules were extended to the conservation of Spin Angular Momenta (SAM) and Orbital Angular Momenta (OAM) when using driving pulses carrying these angular momenta  \cite{Hickstein2015, Fleischer_2014, gauthier2017tunable, Kong2017},  and to collinear schemes with beams of different wavelengths. However, a series of questions remains unresolved: should either Sum Frequency Generation or Difference Frequency Generation be favored (SFG or DFG)? Contradictory results are presented in Refs. \cite{BertrandPRL2011,Heyl2014} on this question. Why does the yield of some diffraction orders decrease when the perturbation intensity increases? Can it be fully accounted for by phase matching arguments, generalizing the conclusion of Ref.  \cite{Heyl2014} to non-perturbative cases? In this work, we address theoretically and experimentally these questions by first analyzing the driving field at focus. We predict in a wide range of perturbation levels the location of the dominant diffraction orders. The toy model proposed is tested in section \ref{QuantumModel} against a full theory of HHG and against experimental data in section \ref{Sec: exp}.  We conclude that, although conservation rules are useful for insights into possible mechanisms, they are insufficient to describe the whole process which, on the contrary is highly dependent on the wavelength scale interferences at focus, in agreement with our toy model. 

\section{Structure of a two beam focus}\label{Sec:analytical}
Like any non linear process, HHG is dramatically influenced by phase matching. In general, the phase matching condition for harmonic q reads
\begin{equation}
\Delta \vec{k}_{XUV}=\vec{k}_q-q\vec{k}_1-\vec{K}
\label{eq:PhaseMatching}
\end{equation}
where $\Delta \vec{k}_{XUV}$ is the phase mismatch that should be minimized for efficient conversion, $\vec{k}_q$ (resp. $\vec{k}_1$) is the wave vector of the harmonic (resp. driving) beam and $\vec{K}$ is grouping the gradients of the intensity dependent atomic phase, which is linked to the HHG process at the atomic level, the Gouy phase of the driving beam, and the phases due to the electron and neutral dispersions in the generating medium. This analysis has been extremely successful for understanding spatio-spectral structures of the emitted harmonics driven by a single beam  \cite{BalcouPRA1997,ConstantPRL1999}, and drives the design of efficient generating schemes (see e.g.  \cite{PaulNature2003, Boutu2011,Rothhardt2014a, Haedrich2015,Heyl2016, Gonzalez2018}). Of particular interest are maps of the modulus of $\Delta \vec{k}+{XUV}$, which inform us about the dominant generating regions of the gas target, while the direction of $\vec{k}_q$ determines the emission direction. This analysis has been generalized to HHG with two  non collinear beams  in Ref.~\onlinecite{Heyl2014}. A geometric additional wave vector was identified, and made responsible for the experimental observation that either DFG or SFG may dominate, depending on the focusing parameters  \cite{Heyl2014} or ionization  \cite{Ellis2017}. We here argue that this ``macroscopic'' approach, which disregards the fine structure of the two beam focus, is only an approximation  limited to weak perturbations. On the contrary, we show that ``mesoscopic'' aspects, at the wavelength scale, play the central part. 
\begin{figure}[!htp]
\centering
\includegraphics[width=0.5\textwidth]{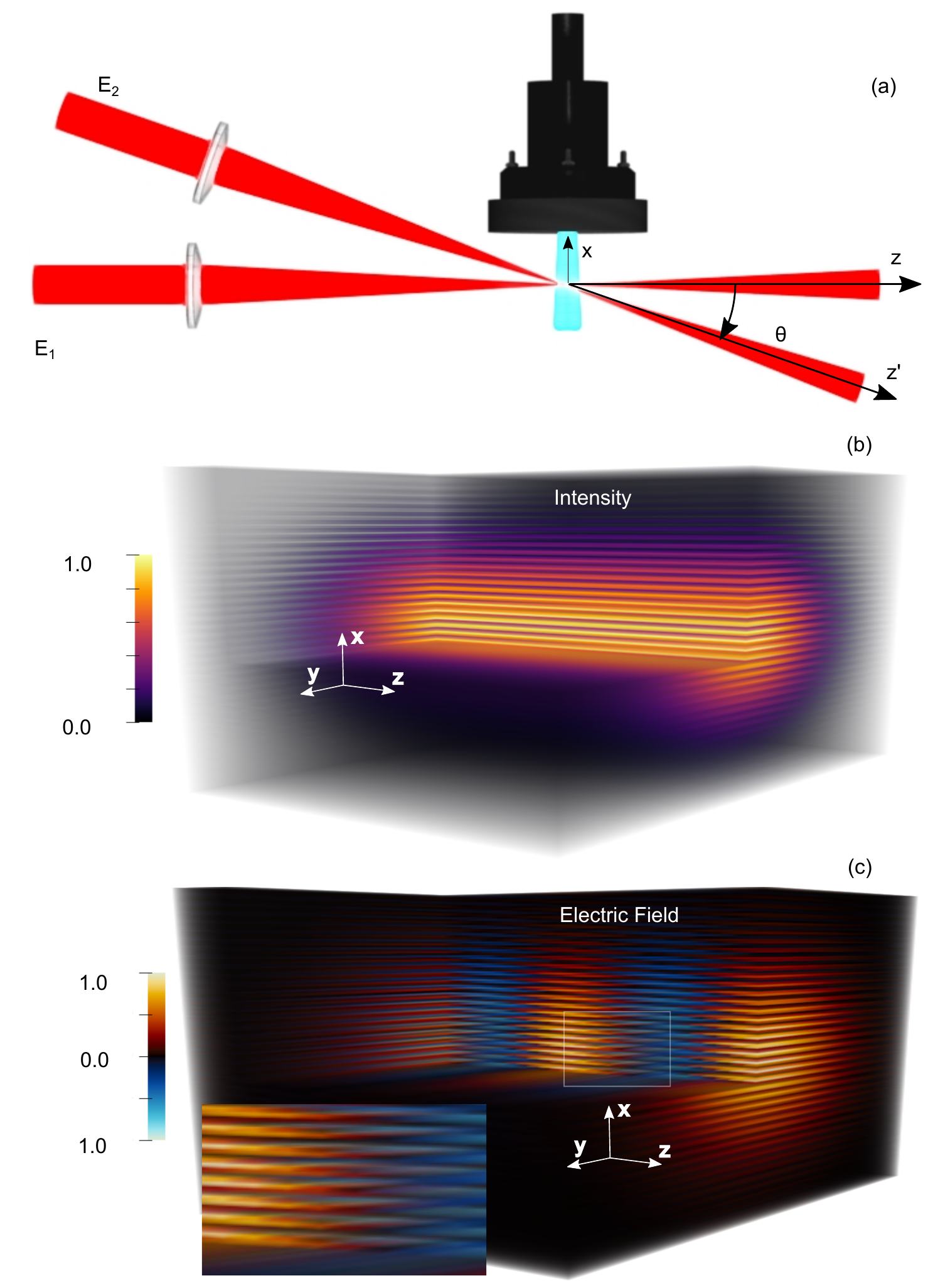}
\caption{(a) Schematic of HHG with two non collinear beams. (b) Intensity map close to focus, when the second beam has an amplitude of 20\% compared to the main beam. (c) Corresponding electric field, with a zoom over the white rectangle plotted in the inset. The first beam is propagating along z, the second z'.     }
\label{Fig:Schematic}
\end{figure}

To illustrate this, we plotted in Fig.\,\ref{Fig:Schematic} the intensity map formed by two Gaussian beams of equal angular frequency $\omega$, linearly polarized along $x$ \footnote{Throughout the article, we consider linear polarizations along y and there is no (weak) longitudinal fields along z}, and wave vectors $\vec{k}_1$ and $\vec{k}_2$ forming an angle $\theta$. The main beam, propagating along the \emph{z}-axis, has an amplitude $E_1$, while the ``perturbative'' beam has the amplitude $E_2=\alpha E_1$. In the focal volume, the intensity forms the grating of planes parallel to (yOz) which is found in optics textbooks. More subtle is the structure of the electric field also plotted in Fig.~\ref{Fig:Schematic}. At a given time and in a given transverse plane, the interference pattern is retrieved. However, careful inspection shows that we can no longer associate planes of equal phases parallel to (yOz) (see the vertical cut along (xOz) where fringes are pointing slightly downwards). Defining a unique wave vector for the field thus becomes impossible and Eq. \eqref{eq:PhaseMatching} should be made local: 
\begin{equation}
\Delta \vec{k}_{XUV}(\vec{r})=\vec{k}_q(\vec{r})-q\vec{k}_s(\vec{r})-\vec{K}(\vec{r}).
\label{eq:PhaseMatching2}
\end{equation}
where $\vec{r}$ is the coordinate vector in the medium and $\vec{k}_s(\vec{r})$ is the local wave vector associated to the sum of the two driving fields. It should be noted that in previous analysis (e.g. Ref.  \cite{BalcouPRA1997,ConstantPRL1999}), this ``local'' form of the phase matching relation was implicitly used for both $\vec{k}_q(\vec{r})$ and $\vec{K}(\vec{r})$. However, the local form of the driving wave vector was mostly discarded: the general case of spatio temporal transient phase matching was considered in detail  \cite{Bahabad2010,Bahabad2011} but only a few specific practical cases of ``guided generation''  were demonstrated, taking into consideration the longitudinal variations of the fundamental wave vector  \cite{zhangnatphysics2007,Sidorenko2010}.

\subsection{Plane continuous waves}
\subsubsection{Analytical signal associated to an interference pattern}
\begin{figure}[!htp]
\centering
\includegraphics[width=0.5\textwidth]{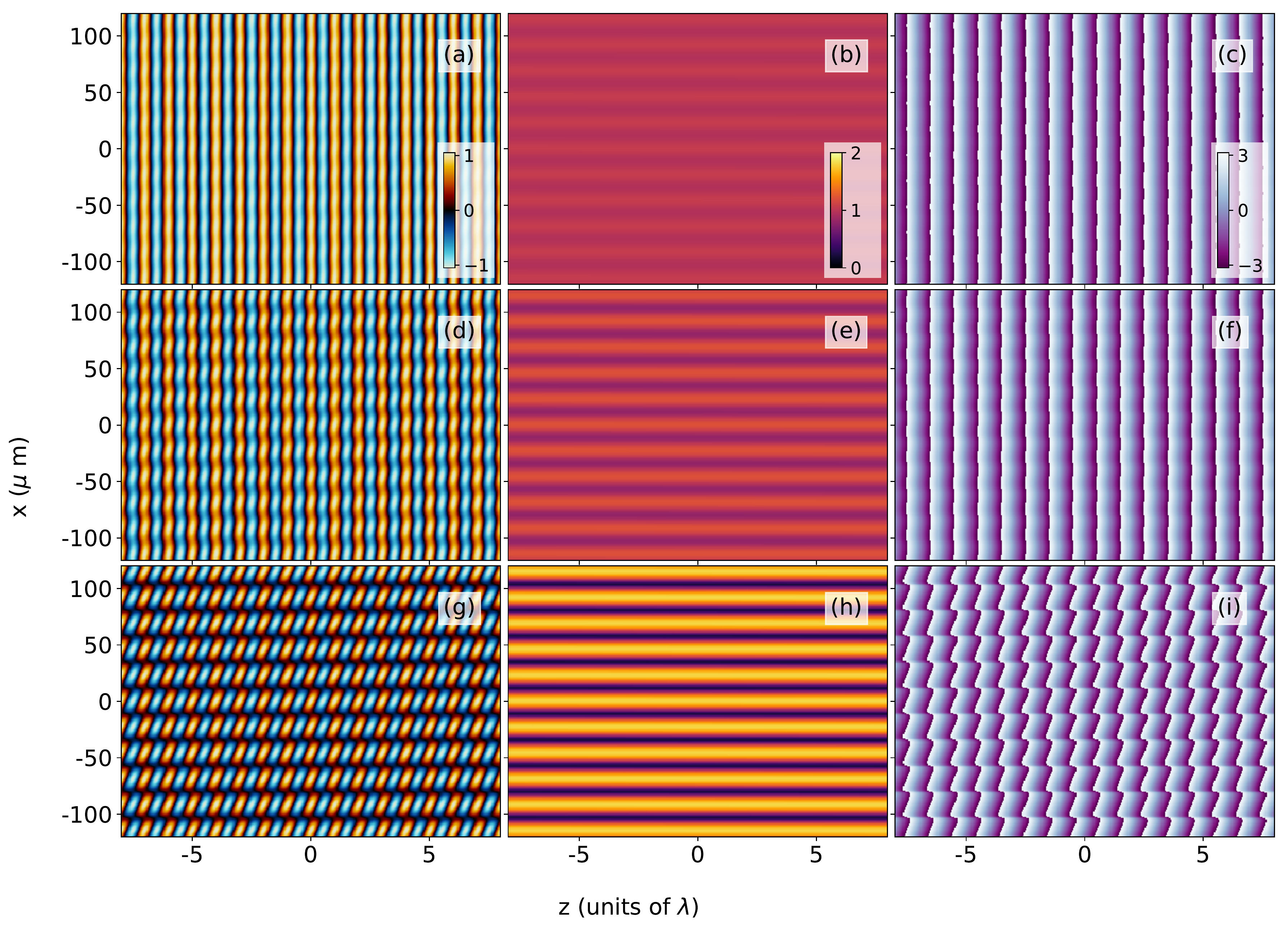}
\caption{(a) (left column) Electric field at focus resulting from the  superposition of two beams making an angle of 2 degrees with relative amplitude of 1:0.05 (top line), 1:0.2 (middle line) and 1:0.8 (bottom line). One beam is horizontal, the second, less intense is coming from top left of the figure. Middle and right columns: modulus and phase of the associated complex analytical signal. The color bars are the same for all columns.  }
\label{Fig:FieldMapsPlane}
\end{figure}

To illustrate pedagogically why this spatial dependence should be introduced, we first consider a field with no envelope, either spatially or temporally. The total field in the jet reads 
\begin{align}
E_{s}&=\mathcal{R} \left[E_1 e^{i\omega t -i \vec{k}_1\cdot \vec{r}}+\alpha E_1 e^{i\omega t -i \vec{k}_2\cdot \vec{r}}\right]\\
&=E_1 \mathcal{R}\left[ e^{i\omega t -i \vec{k}_1\cdot \vec{r}}\left(1+\alpha e^{ -i \vec{\Delta k}\cdot \vec{r}}\right)\right]
\label{eq:sumGeneral}
\end{align}
where the expression in brackets is the analytical signal, denoted $\tilde{E}_s$, associated to the real field $E_{s}$ and  $\vec{\Delta k}=\vec{k}_2-\vec{k}_1$. 
Cuts of the electric field  through the \emph{(xOz)} plane are displayed in Fig. \ref{Fig:FieldMapsPlane} (a,d,g) for three amplitudes of the perturbation ($\alpha=$5\%, 20\% and 80\%).
The perfect plane waves iso-amplitude curves along the Ox direction are progressively destroyed as the perturbation increases, finally yielding a checkerboard pattern of high intensity zones, where HHG will occur. This pattern can be decomposed in the form of an analytical signal associated to the real field, yielding the amplitude and phase maps displayed in Fig. \ref{Fig:FieldMapsPlane}, 2nd and 3rd column. We retrieve the usual intensity map of an interference pattern with increasing contrast as $\alpha$ increases. More interesting for our topic are the phase maps. They show a small modulation along the \emph{x}-direction, which progressively transforms into a sawtooth pattern. Interestingly, the zones of high amplitude of the analytical signal are associated to a stationary and  downward tilted wavefront (see e.g. Fig. \ref{Fig:FieldMapsPlane}(h-i)). We can thus anticipate an HHG emission favored downstream, in the direction of SFG. 

\subsubsection{Local wave vector}
\begin{figure}[!htp]
\centering
\includegraphics[width=0.5\textwidth]{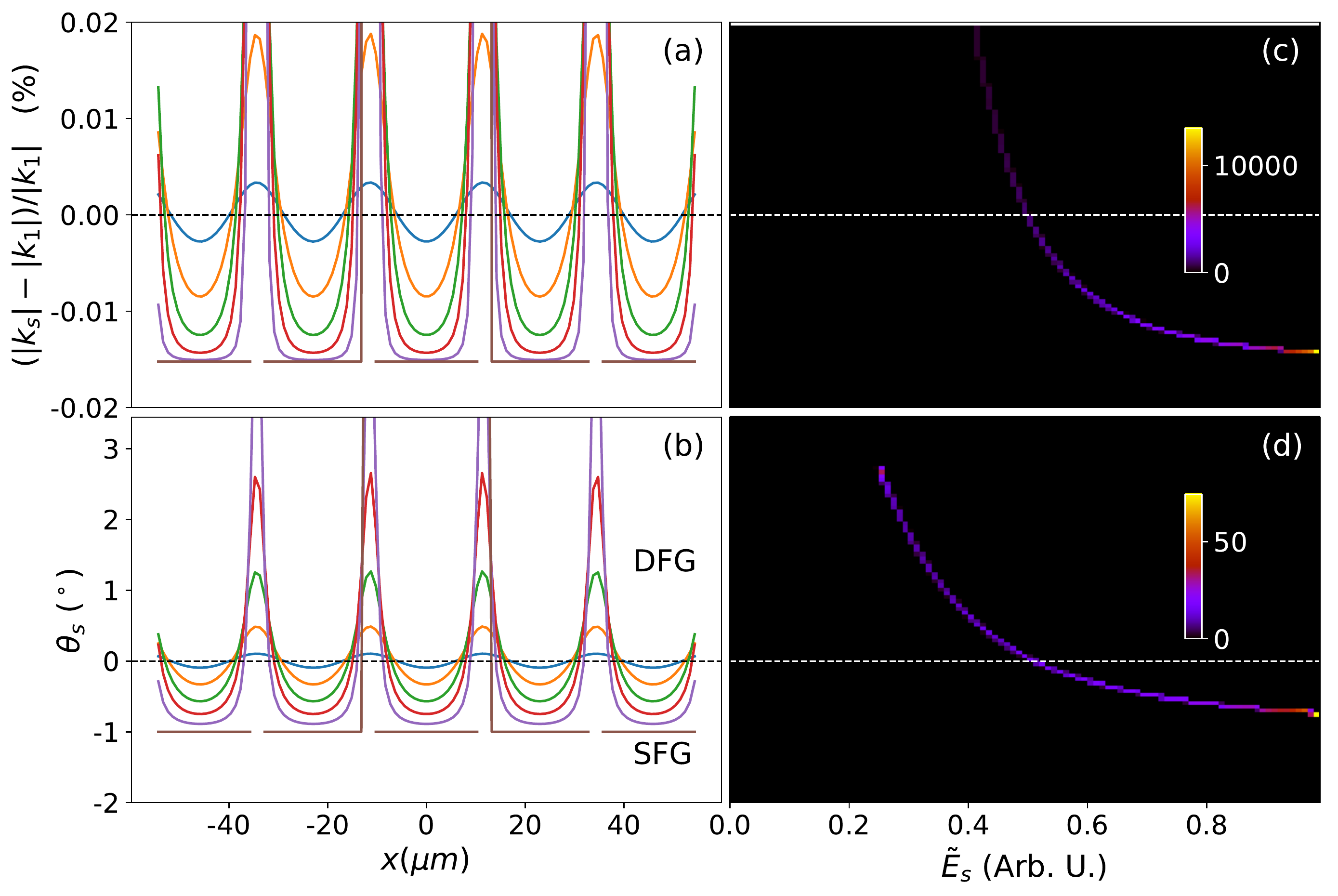}
\caption{Wave vector modulus (a) and angle with respect to horizontal (b) across the transverse direction (x) for different ratios of the two fields: 1:0.05 (blue), 1:0.2 (orange), 1:0.4 (green), 1:0.6 (red), 1:0.8 (purple) and 1:1 (brown). The data corresponding to amplitudes of the analytical signal below 10\% of its maximum has been discarded. (c-d) Histograms of the distributions of $\frac{|k_s|-|k_1|}{|k_1|}$ and $\theta_s$ for the case of a $\alpha=0.6$. Each vertical cut is a histogram for a given field amplitude $\tilde{E}$ giving finally these 2D histograms. The amplitude of the analytical signal $\tilde{E}_s$ has been normalized to 1.   }
\label{Fig:FieldMaps}
\end{figure}

 To get more insight, we now derive an expression of the local wave vector.
We reorder the last part of Eq. \eqref{eq:sumGeneral}  as
\begin{equation}
1+\alpha e^{ -i \vec{\Delta k}\cdot \vec{r}}=f_m(\alpha,\vec{\Delta k}\cdot \vec{r}) e^{-i\varphi(\alpha,\vec{\Delta k}\cdot \vec{r})}
\label{eq:Localwavevector}
\end{equation}
with
\begin{align}
f_m(\alpha,\vec{\Delta k}\cdot \vec{r})&=\sqrt{1+\alpha^2+2\alpha\cos\left( \vec{\Delta k}\cdot \vec{r}\right)}\label{AnalForm1}\\
\varphi(\alpha,\vec{\Delta k\cdot \vec{r}})&=\arctan{\frac{\alpha \sin\left( \vec{\Delta k}\cdot \vec{r}\right)}{1+\alpha\cos\left( \vec{\Delta k}\cdot \vec{r}\right)}}
\label{AnalForm2}
\end{align}
We thus get the analytical representation of the composite field 
\begin{equation}
\tilde{E}_{s}=E_1 f_m(\alpha,\vec{\Delta k}\cdot \vec{r})\cdot e^{i\omega t -i \vec{k}_1\cdot \vec{r}-i\varphi(\alpha,\vec{\Delta k\cdot \vec{r}})}
\label{eq:AnalSignal0}
\end{equation}
The local wave vector of the sum of the two fields, denoted $\vec{k}_s(\vec{r})$, is given by the spatial gradient of the phase. Some standard algebra leads to \footnote{The main formulas of this section are derived in the Appendix}
\begin{equation}
\vec{k}_s(\vec{r})=\vec{k}_1+\vec{\nabla}\varphi=\vec{k}_1+\frac{\alpha \left(\alpha+\cos( \vec{\Delta k}\cdot \vec{r})\right)}{1+\alpha^2+2\alpha\cos( \vec{\Delta k}\cdot \vec{r})}\vec{\Delta k}.
\label{eq:wavevector}
\end{equation}
In the (xOz) plane, we denote 
\begin{align}
\vec{k}_s &= \begin{bmatrix}
k_s \sin \theta_{s}\\
k_s \cos\theta_{s}
         \end{bmatrix}.
\label{eq:def_ks}
\end{align}
As expected, we find that the initial wave vector is perturbed by an almost orthogonal contribution (along $\vec{\Delta k}$, last part of Eq. \eqref{eq:wavevector}), with an amplitude that strongly depends on the location in the medium. The relative amplitude of this ``active grating'' contribution and its angle with respect to the horizontal z-axis are displayed in Fig. \ref{Fig:FieldMaps}(a-b) for a series of $\alpha$ against the transverse dimension. At very low perturbation levels, both the wave vector amplitude and angle vary sinusoidally against \emph{x}. The excursion is perfectly up/down symmetric. However, as soon as the perturbation increases, the spectral content gets richer, finally converging to constant values for $\alpha=1$, $\theta_{s}=\theta/2$ for the angle, and $|\Delta k_1|/|k_1|\simeq -1.5\times 10^{-4}$. The angle is simply the bissector of the two beams, as expected for two equally intense interfering beams.
In the following, we first consider this general expression giving the full map of the wave vector in the cases of weak and strong perturbations, before focusing on vertical lines where it is stationary, e.g. at x=0.

\subsubsection{Strong perturbation case}
In quite a few current schemes, equally strong fields are asked for (e.g. Ref.  \cite{Hickstein2015}). To get the wave vector in this case, we set $\alpha=1-\beta$, with $\beta\ll 1$ in Eq. \eqref{eq:wavevector}. We then get
\begin{equation}
\vec{k}_s(\vec{r})=\vec{k}_1+\frac{\vec{\Delta k}}{2}-\frac{\beta}{1+\cos( \vec{\Delta k}\cdot \vec{r})}\frac{\vec{\Delta k}}{2}.
\label{eq:StrongPertub}
\end{equation}
As a first approximation, the wave vector is determined by the bisector of the two beams, as it could be intuited: 
\begin{equation}
\vec{k}_s(\vec{r})=\vec{k}_1+\vec{\Delta k}/2 \;\; \text{(for $\alpha = 1$,\quad $\beta=0$)}
\label{eq:WaveVectorAlpha1}
\end{equation}
It is independent of the location in the medium.  With this expression, some algebraic manipulations give the following expressions for the modulus and angle of the local wave vector:
\begin{align}
\frac{|k_s|-|k_1|}{|k_1|}&\simeq -\frac{1}{2}\sin^2 \frac{\theta}{2}\label{Modulus_alpha1}\\
\theta_{s}&=\theta/2\label{Angle_Alpha1}
\end{align}
which correspond to the values found in Fig. \ref{Fig:FieldMaps} with $\theta=-2^{\circ}$. In case of perfect symmetry between the two beams, we thus have a homogeneous wavefront throughout the medium. It is tilted with respect to the horizontal axis by half the angle between the two beams. We thus expect all harmonics to be emitted along this direction. However, the magnitude of the wave vector will be modified significantly and should alter the phase matching conditions. This last effect had not been considered thus far. 

In case of two beams non perfectly symmetric, which could be the result of imperfect matching of the energies, spatial or temporal overlaps, the last term in Eq. \eqref{eq:StrongPertub} should be taken into account. Now the wave vector is non homogeneous spatially. We thus predict  a dispersion of the amplitude and emission direction of the harmonics along several diffraction orders. Also, we note that formula \eqref{eq:StrongPertub} is asymmetric with respect to the bisector. We thus anticipate that positive and negative orders will not be equally strong. This could explain the results observed for instance in Ref.   \cite{Hickstein2015} (Sup. mat.).

\subsubsection{Weak perturbation case}

The second interesting case, which was investigated in  \cite{BertrandPRL2011,Heyl2014}, is the perturbative case, $\alpha \ll 1$. Here
\begin{equation}
\vec{k}_s(\vec{r})=\vec{k}_1+\alpha\vec{\Delta k}\cos\left(\vec{\Delta k}\cdot \vec{r}\right) \;\; \text{(for $\alpha \ll 1$)}
\label{eq:LocalLowAlpha}
\end{equation}  
with the corresponding modulus and angle of the wave vector
\begin{align}
\frac{|k_s|-|k_1|}{|k_1|}&\simeq -2\alpha\sin^2\frac{\theta}{2}\cos( \vec{\Delta k}\cdot \vec{r})\label{Modulus_alpha0}\\
\theta_{s}&\simeq\alpha\sin\theta\cos( \vec{\Delta k}\cdot \vec{r})\label{Angle_Alpha0}
\end{align}
The wave vector is modulated in both direction and modulus sinusoidally, justifying the ``local form'' in Eq. \eqref{eq:PhaseMatching2}. HHG being highly non linear, the intensity modulation will transfer to a phase spatial modulation affecting phase matching, just as the modulation in amplitude and direction of the wave vector will. Importantly, the modulation is here perfectly symmetrical along the transverse direction. Identical generating volumes thus  have wave vectors pointing upwards and downwards, making no difference between the SFG and DFG amplitudes. We can thus anticipate equally strong positive and negative diffraction orders in the far field. 

\subsubsection{General case and discussion}
Such analytical formula, which are valid everywhere in the medium, cannot be derived easily for intermediate perturbation levels. However, we can still get insights into the consequences of the modifications of the wave vector by considering its stationary values in the generating volume. Getting back to the microscopic ``atomic'' level, HHG appears as a three step process. Close to an extremum of the driving field, a valence electron is tunnel ionized (first step), generating an electronic wave packet (EWP). The EWP, launched in the continuum, is driven away from its ionic core before being pulled back when the field changes sign, about a quarter of a period later (second step: excursion in the continuum). Finally, under specific initial conditions, the EWP may recollide with the ionic core and recombine, emitting its excess of energy as an XUV photon (third step: recombination).  The first step is a highly non linear effect. Only places where the field is strong will significantly contribute to the far field amplitude. Moreover, simple computations show that the maximum energy of an emitted photon reads $E_{cutoff}=I_p+3.2 U_p$ where $I_p$ is the ionizing potential of the atom and $U_p$ the ponderomotive potential of the field, which scales as $U_p\propto \tilde{E}^2 \lambda_1^2$. The highest XUV photons can thus only be generated at locations in the medium where the field is strong enough to be authorized by the cutoff law. On the contrary, lower energy photons,  lying in the so-called plateau of the spectrum, can be generated by these strong fields, but also weaker fields. 

These considerations motivated the plot of the 2D histograms of Fig. \ref{Fig:FieldMaps}. They display the joint probability of given ($|\vec{k}_s|$, $\tilde{E}_s$) and ($\theta_{s}$, $\tilde{E}_s$). On these maps, the stronger the color, the more probable the couple of values. In this case, we observe that the highest harmonics, which can only be generated for the highest field amplitudes $\tilde{E}_s$, will be generated with a stationary value of the amplitude and inclination angle of the driving wave vector.  On the contrary, for the plateau harmonics, which can be generated for many values of $\tilde{E}_s$, there will be a spreading of the angles of emission, which is all the wider as the photon energy is low. To be more specific, it appears in Fig. \ref{Fig:FieldMaps} (a-b) that the wave vector is stationary at $x=0$. The wave vector there reads
\begin{equation}
\vec{k}_s(\vec{r}=0)=\vec{k}_1+\frac{\alpha}{1+\alpha} \vec{\Delta k}
\label{eq:GeneralWaveVectorStationary}
\end{equation} 
It corresponds to a magnitude and angle 
\begin{align}
|\vec{k}_s(\vec{r}=0)|=&|\vec{k}_1|\sqrt{1-\frac{4\alpha}{(1+\alpha)^2}\sin^2\left( \frac{\theta}{2}\right)}\label{Modulus_alphaGeneral}\\
\theta_{s}&=\arctan\frac{\alpha\sin\theta}{1+\alpha\cos\theta}.\label{Angle_General}
\end{align}
The angle $\theta_{s}$ is not simply the average of the two wave vectors as it could be intuitively guessed. However, it goes smoothly from a linear behavior to its final value $\theta/2$ as $\alpha$ increases. As for the magnitude, considering small angles between the two beams we get the expression
\begin{equation}
\frac{|k_s(\vec{r}=0)|-|k_1|}{|k_1|}\simeq -\frac{2\alpha}{(1+\alpha)^2}\sin^2\left( \frac{\theta}{2}\right)
\label{eq:GeneralWaveVectorStationaryApprox}
\end{equation} 

\begin{figure}[!htp]
\centering
\includegraphics[width=0.5\textwidth]{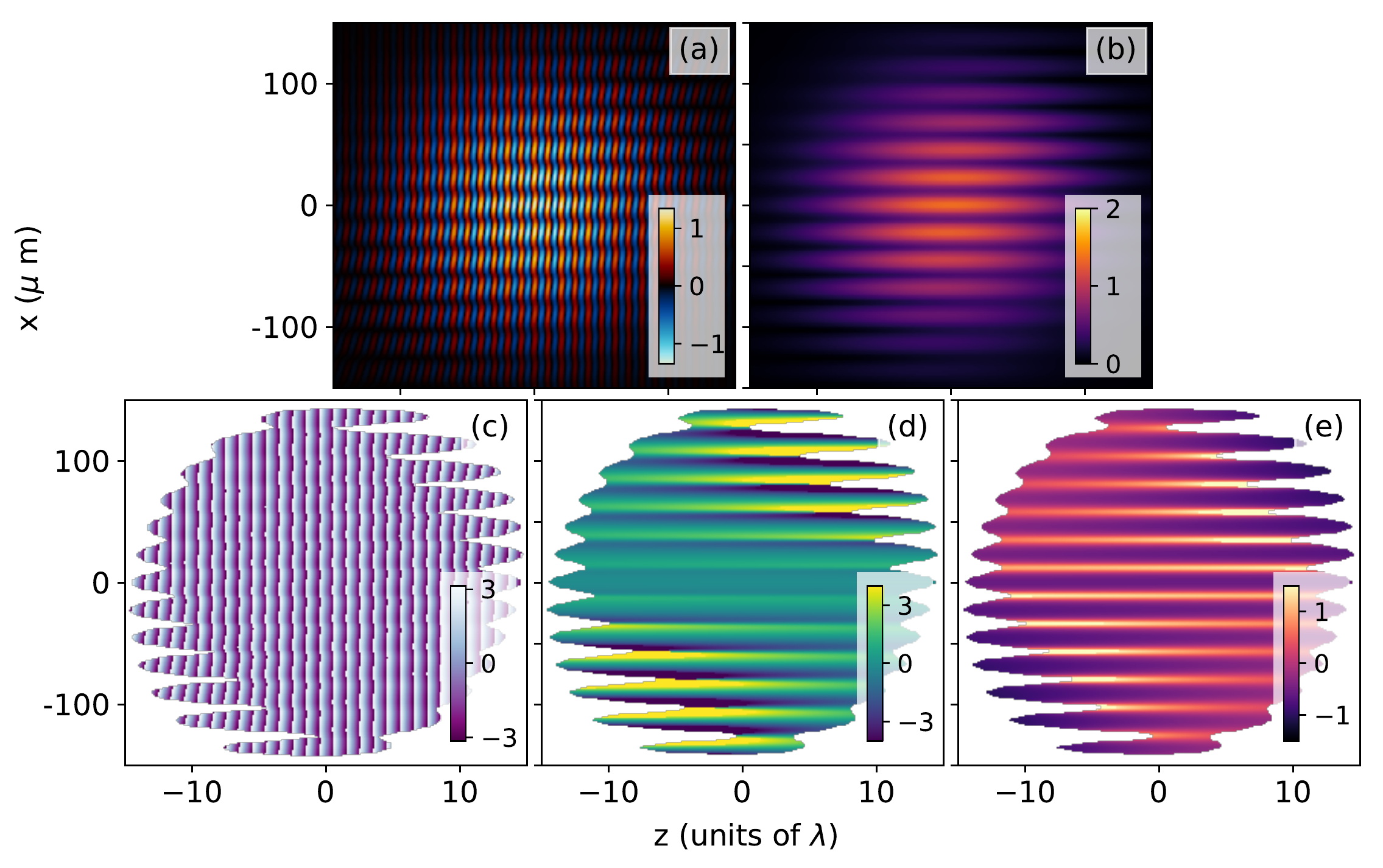}
\caption{(a) Electric field at focus resulting from the  superposition of two beams making an angle of 2 degrees with relative amplitude of 1:0.4. The two waists are 100$\mu m$, durations of 30\,fs, $\lambda$=800\,nm. (b) Amplitude and (c) phase of the associated analytical signal. (d) Relative amplitude ($|k_s(\vec{r})|-k_1$ in $10^4$ m$^{-1}$) and (e) inclination angle (in degrees) of the local wave vector. For maps (c-e) all data corresponding to magnitudes of the analytical signal below 10\% of its maximum have been discarded.  }
\label{Fig:FieldMapsGaussian}
\end{figure}

It should be noted that this additional phase mismatch modulation due to the ``active grating'' is comparable in amplitude to usual phase matching factors. To give a few orders of magnitude, for an angle of 2$^{\circ}$ and 10\% perturbation in amplitude, the excursion of the wave vector magnitude is $\simeq \pm q\times 0.5 rad/mm $, where q is the harmonic number. It would correspond to a coherence length of 1/q\,mm if alone. As a comparison, the new term found in ref.  \cite{Heyl2014} has a magnitude of about $0.2 \;rad/mm $, and in standard conditions coherence lengths are found on the mm or fraction of mm scale, with the different terms precisely of the order of a fraction of radian per mm (see. e.g.  \cite{RoosPRA1999, KhuongPhD}). The term we identified here is thus far from negligible, and could even become dominant.  

To summarize this first analysis, we identified that the local wave vector of the driving field is strongly modified in both direction and amplitude by the presence of a second beam, which should dramatically affect the global HHG yield. 

\subsection{Gaussian spatial and temporal profiles}
To get a more realistic situation, we now include the spatial and temporal profiles of the beams in the simulations. The conclusions drawn above still hold ``locally'', for a given ratio of the fields. However, even in a  generating medium of only a few tens of microns long, with waists of the beams of about 100$\mu m$, significant variations of the relative amplitude of the beam occur in the medium. This is made apparent in Fig. \ref{Fig:FieldMapsGaussian}. As expected, the electric field, magnitude and phase of the associated analytical signal show patterns very similar to those of Fig. \ref{Fig:FieldMapsPlane}, but for overall spatial (along \emph{x}) and temporal (along\emph{ z}) envelopes.  
The most striking feature appears in Fig. \ref{Fig:FieldMapsGaussian} (d-e). As in Fig. \ref{Fig:FieldMaps}, the main trend is a transverse modulation of the magnitude and angle of the local wave vector. However, there is a left/right and up/down asymmetry appearing. It is naturally due to the varying relative amplitudes of the beams at different locations in the medium. 
\begin{figure}[!htp]
\centering
\includegraphics[width=0.5\textwidth]{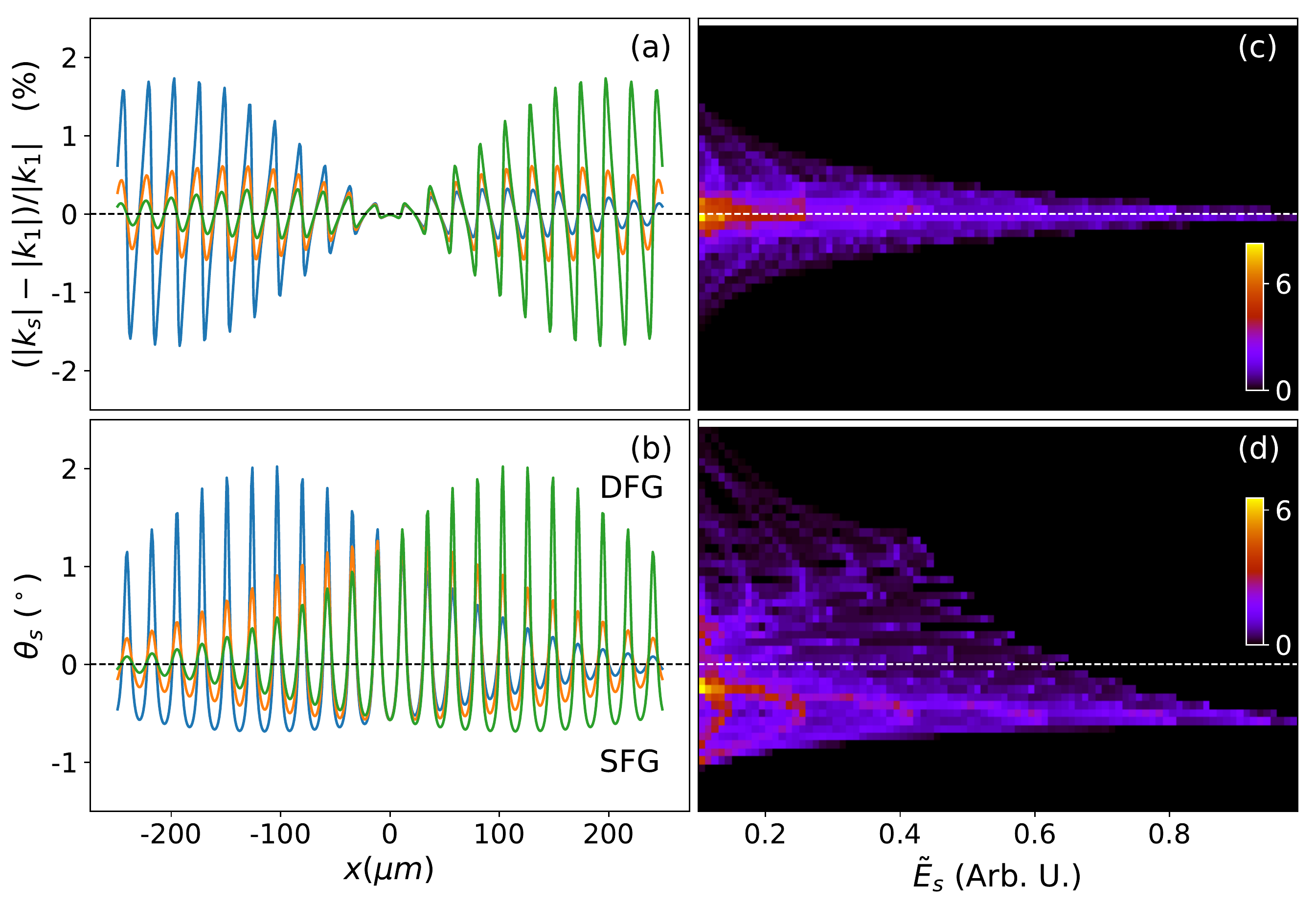}
\caption{ (a-b) Cuts through the maps displayed in Fig. \ref{Fig:FieldMapsGaussian}(d-e) at z=0 (orange), z=5\;$\lambda$ (Blue) and z=-5\;$\lambda$ (green) along the transverse direction. For $\theta_{s}$, negative inclination angle means propagation towards the bottom right of the figure, i.e. SFG processes with our convention. (c-d) Two dimensional histograms corresponding to Fig. \ref{Fig:FieldMapsGaussian}(d-e), like in Fig. \ref{Fig:FieldMaps}.  }
\label{Fig:HistoGaussian}
\end{figure}
To quantify this asymmetry, we plotted in Fig. \ref{Fig:HistoGaussian}(a-b) three lineouts of these maps, along with their 2D histograms vs. the amplitude of the field. While the central cut is top/bottom symmetric (orange), the two others are mirror images. For instance, for the cut at $-5\;\lambda$ (green), the beam should interfere more at positive x (the tilted beam comes from the top). Accordingly, the wave vector is slightly modulated and mainly points towards the horizontal at negative x. It should be noted that the effect is large, reaching 1.5\% of excursion for $|\Delta k|$ and more than 2$^{\circ}$ for $\theta_{s}$. The 2D histograms show that the distribution of $|\Delta k|$ remains limited to a few permil, depending on the value of $\tilde{E}_s$ considered. The angle shows an ``Eiffel tower'' like distribution which points towards the stationary angle $\theta\cdot\alpha/(1+\alpha)$, given by Eq. \eqref{Angle_General}. It is interesting to note that the introduction of the envelopes of the field changed the general histograms drastically: whereas a bijection is observed between $\tilde{E}_s$, $|\vec{k}_s(\vec{r})|$ and $\theta_{s}$ for plane waves, here an almost  constant value is obtained for  $|\vec{k}_s(\vec{r})|$ and $\theta_{s}$. The only varying parameter is the distribution about this mean value that decreases as $\tilde{E}_s$ increases. 
\begin{figure}[!htp]
\centering
\includegraphics[width=0.5\textwidth]{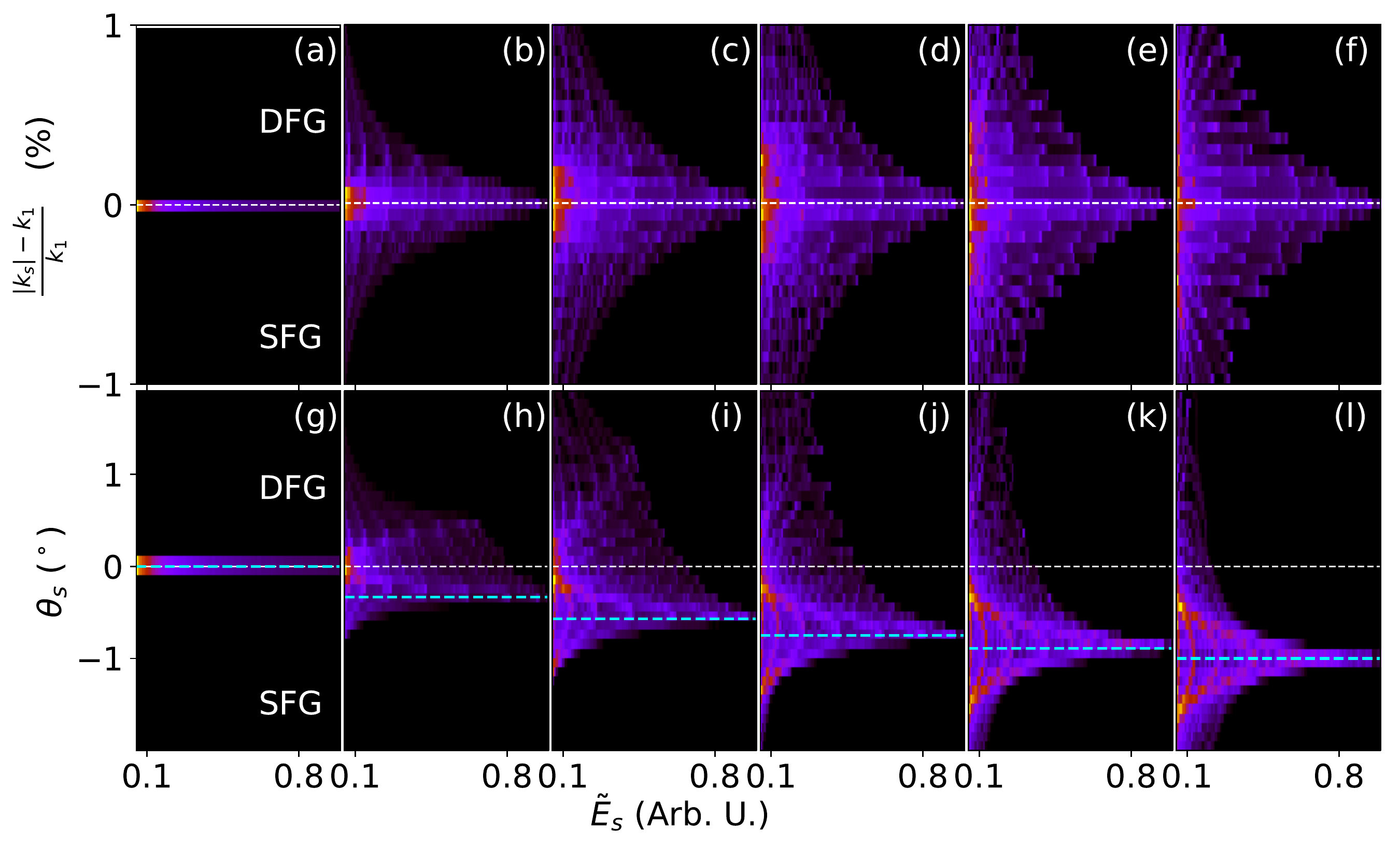}
\caption{Same histograms as in Fig. \ref{Fig:HistoGaussian} for different ratios ($\alpha=0, 0.2, 0.4, 0.6 ,0.8,1.0$ from left to right). The light blue dashed line is at $\theta\cdot\alpha/(1+\alpha)$.    }
\label{Fig:FieldHisto}
\end{figure}

In Fig. \ref{Fig:FieldHisto}, we plotted the same histograms for a series of $\alpha$'s. We note that the distribution of $|\vec{k}_s|$ increases dramatically with $\alpha$. We will thus have good phase matching in an ever decreasing volume as $\alpha$ increases. At the same time, the ``Eiffel tower'' shape converges to a perfectly symmetric shape about the half angle between the beams, which is expected, but with an ever increasing base. We can thus infer, as a first approximation, that a regular blazed active phase grating is created in the medium, with a phase varying like 
\begin{equation}
\varphi_{grating}\simeq \frac{\alpha}{1+\alpha}\cdot\theta \times k_1x.
\label{eq:gratingPhase}
\end{equation} 
This rough approximation of a blazed phase grating will become less and less exact as $\alpha$ increases for harmonics of the plateau while remaining valid for cutoff harmonics.

\section{HHG by a field showing a blazed periodical structure}\label{Sec:blazedGrating}
Having identified that two interfering beams create a blazed phase grating and a symmetric amplitude grating in the focus area, we now investigate its consequence on the HHG process. We will pay specific attention to the symmetries of the SFG and DFG processes. As a very rough model of HHG, we may simply consider the sum of the contributions over a volume 
\begin{equation}
E_q(\vec{r}_{out})=\iiint_V e_q(\vec{r})e^{i\Delta \vec{k}_{XUV}(\vec{r})\cdot\vec{r}}d^3\vec{r}
\label{eq:HHGBasics}
\end{equation}
where $E_q$ is the macroscopic electric field of harmonic q at location $\vec{r}_{out}$ at the exit of the generating medium, $e_q(\vec{r})$ is the microscopic response at location $\vec{r}$ inside the medium and $\Delta \vec{k}_{XUV}(\vec{r})=q\vec{k}_s(\vec{r})-\vec{k}_q(\vec{r})$ is the ``local'' phase mismatch. The integration is carried out over the generating gas volume $V$, which, for the sake of simplicity, we consider to be infinitely small along the z direction. $e_q(\vec{r})$ is highly non linear with the local amplitude of the field. The modulation of the amplitude of the driving field will thus  create a series of generating slits, forming a transverse amplitude grating. Importantly, this grating is perfectly symmetric against the transverse axis x at $z=0$. Through the intensity dependent response of the atoms to strong fields ($\phi_{at}\propto I$), it is also a phase grating, which is also x-wise symmetric at $z=0$. Naturally, off focus, an asymmetry appears versus the x-axis, which will be opposite at symmetrical locations upstream and downstream from the focal spot. Considering a gas jet located at $z=0$, these two effects thus have an overall symmetry against x.

  Of more interest here is the exponential term. According to the previous section (Eq. \eqref{eq:gratingPhase}) the phase is approximately linear with x. We thus have a ``blazed'' phase grating, asymmetric with respect to x, which gets superimposed on the pattern discussed above. This is true everywhere in the medium, at focus, upstream and downstream: the dominant blaze angle always has the same sign.
To get a toy model based on these considerations, we consider a transverse grating made of a series of Gaussian gates, labeled from $-n_0$ to $n_0$, with a Gaussian envelope, and the linear phase $\varphi(x)=q\varphi_0\frac{x}{\Lambda}$ for harmonic q. In agreement with Eq. \eqref{eq:gratingPhase} we set 
\begin{figure}[!htp]
\centering
\includegraphics[width=0.5\textwidth]{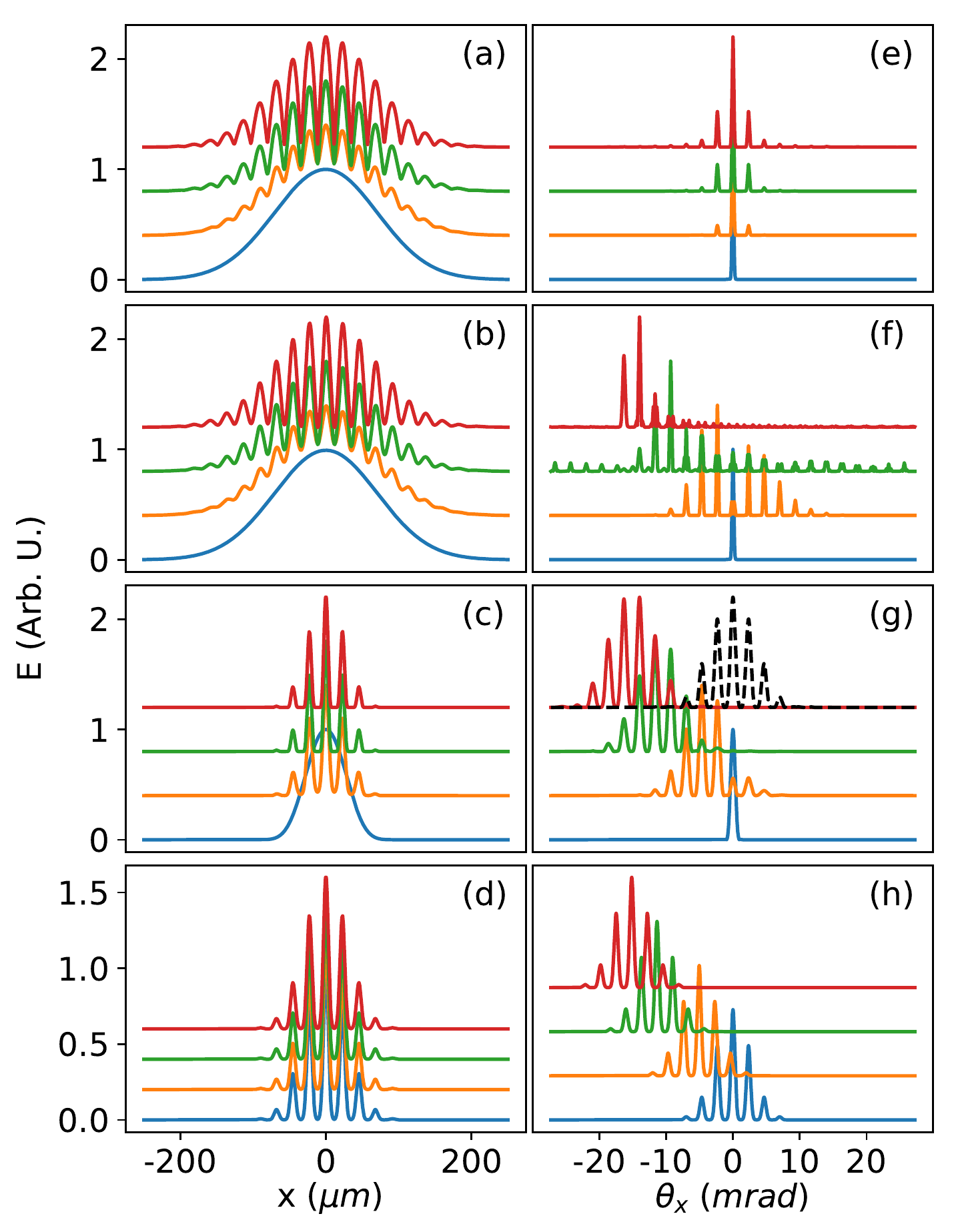}
\caption{Intensities at $z=0$ (a-d) and in the far field (e-h)  for $\alpha=0$ (blue), $\alpha=0.2$ (orange), $\alpha=0.6$ (green) and $\alpha=1.0$ (red). We consider H15. For the sake of visibility the curves are all normalized to 1 and shifted respectively by 0, 0.4, 0.8 and 1.2. (a) and (b) correspond to cuts analogous to the orange curve in Fig. \ref{Fig:HistoGaussian} (b). (c) is the amplitude of the ionization rate computed with the ADK model. (d) is a toy model corresponding to a series of gaussian amplitude with periodicity $\Lambda=1/\Delta k$, of width $\Lambda/5$ with a gaussian global envelope of width 60$\mu m$. The phase is here set according to Eq. \eqref{eq:gratingPhase}. (e-h) Amplitudes of the Fourier transforms of (a-d). In (e), the phase of the signal is set to zero; in (f-g) the phases computed in Fig. \ref{Fig:HistoGaussian} are plugged; in (h) we took the analytical formula of Eq. \eqref{eq:phi_0}. The results are offset by the same quantities and the color code is the same on all plots. The distance is z=0.3 m. In (g) the black dotted line corresponds to the same as the red, but setting the phase to 0.   }
\label{Fig:FigCuts}
\end{figure}
\begin{equation}
\varphi_0=\frac{\alpha}{1+\alpha}\times 2\pi
\label{eq:phi_0}
\end{equation}
The harmonic field at the exit of the generating medium can thus be written:
\begin{equation}
\begin{split}
E_q(\vec{r}_{out})&=\left[G{\left(\frac{x}{\delta x}\right)}\times e^{iq\varphi_0\frac{x}{\Lambda}} \right]*\\
&\quad\quad\left[\sum_{n=-n_0}^{n_0}\delta\left(\frac{x}{\Lambda}-n\right) \times G\left(\frac{x}{\Delta x}\right)\right]
\label{eq:GratingOutput}
\end{split}
\end{equation}
where G is the gaussian function $G(x)=e^{-x^2}$, $\Lambda$ the periodicity of the grating ($\Lambda=2\pi/|k_1-k_2|=\lambda/\sin\theta$ is the periodicity of the intensity of the sum of the fields), $\delta x$ the width of each slit and $\Delta x$ the width of the envelope. In the Fraunhoffer diffraction regime,  at distance $z_{ff}$ from the medium, the amplitude reads 
\begin{multline}
E_q(\vec{r}_{ff})\propto \left[ G\left(\pi\delta x (u-q\frac{\varphi_0}{2\pi\Lambda})\right)\times\sum_{n=-n_0}^{n_0}e^{-i 2\pi n \Lambda u}\right] *\\
  G\left(\pi\Delta x u\right)
\end{multline}
\begin{multline}
\phantom{{}E_q(\vec{r}_{ff}){}}\propto 
\left[ G\left(\pi\delta x (u-q\frac{\varphi_0}{2\pi\Lambda})\right)\times \right.\\\left. e^{-i\pi\Lambda u} \frac{\sin(2\pi n_0\Lambda u)}{\sin(\pi\Lambda u)}\right] *
  G\left(\pi\Delta x u\right)
\label{eq:fieldFarField}
\end{multline}
 where $u=x'/\lambda_q z_{ff}$ with $x'$ labeling the vertical coordinate at the observation plane and $\lambda_q=\lambda/q$ the wavelength of harmonic q. The first Gaussian has an envelope of width 
\begin{equation}
w_q=\frac{\lambda_q z_{ff}}{\pi\delta x},
\label{eq:widthLargeGaussian}
\end{equation} offset by 
\begin{equation}
x_c'=\frac{\lambda}{\Lambda}\cdot\frac{\varphi_0}{2\pi}\cdot z_{ff}.
\label{eq:OffsetOrders}
\end{equation} 
It is multiplied by a function that converges to a series of Dirac peaks as $n_0$ increases. The location of the p$^{th}$ Dirac peak is 
\begin{equation}
x'(q,p)=p\cdot\frac{\lambda}{q\Lambda}\cdot  z_{ff} =p\cdot\frac{\left|\vec{k}_2-\vec{k}_1\right|}{2\pi q}\cdot  z_{ff} 
\label{eq:LocationOrders}
\end{equation} 
The intensity profile in the far field thus appears as a series of peaks, equally spaced  by $x'(q,1)$, centered around $x'_c$, with an amplitude decreasing along a large Gaussian of width $w_q$. The peaks correspond to the diffraction orders. The last convolution only transforms the series of Dirac peaks into physical Gaussian finite functions of small width. The last expression in Eq. \eqref{eq:LocationOrders} was interpreted as a conservation law of momentum during a non linear process  \cite{BertrandPRL2011,Heyl2014,Hickstein2015}: the wave vector of the outgoing photon points in the directions corresponding to that of the driving field plus an integer number of the difference of the wave vectors of the two interfering fields. In other words, for a given harmonic q, a photon picture may be associated to each diffraction order labeled $p$, which corresponds to the absorption (or stimulated emission) of q-p photons of the first beam and p photons from the second beam. However, it should be noted that this ``multiphoton'' picture has nothing to do with being in a perturbative regime and is much more general. It here appears as a consequence of the quasi periodicity of the sum of the two fields in the transverse direction.  

The main point here is that the dominant diffraction order is ruled by the grating depth $\varphi_0$ and is independent of the harmonic order q. q only enters into the spacing of the comb of diffraction orders. This is confirmed in the plots displayed in Fig. \ref{Fig:FigCuts}. Taking a cut at $z=0$, whatever $\alpha \neq 0$, if the phase is not taken into account we get a series of peaks centered about zero (Fig. \ref{Fig:FigCuts}(a,e)). If the phase is considered, the peaks are all the more offset as $\alpha$ increases (Fig. \ref{Fig:FigCuts}(b,f)). HHG being a highly non linear process, especially through the field-driven tunnel ionization constituting the first step of the process, the slits should be fewer and thinner than the oscillations of the electric field. As a very rough approximation of this effect, the HHG signal can be estimated proportional to the tunnel ionization rate at any time. We estimated it using the Amonosov-Delone-Krainov (ADK) formula  \cite{Tong2005} (Fig. \ref{Fig:FigCuts}(c,g)), keeping in mind that it is rigorously valid only for continuous fields. We get only a few half periods ($\simeq 5$) contributing, with widths of a fraction of the field's half period. The consequence in the far field is a smoother profile of the harmonic orders, which still shows a shift towards SFG. This behavior is very well reproduced by the toy model  exposed above (Fig. \ref{Fig:FigCuts}(d)): Eq. \eqref{eq:GratingOutput}, Fig. \ref{Fig:FigCuts}(h): its Fourier transform), enlightening the role of the modulation depth of the phase in the relative intensities of the diffraction orders. 

\section{Full quantum model of HHG in an active grating}\label{QuantumModel}
To be more quantitative, we performed full numerical simulations based on the solution of the nonadiabatic, three-dimensional (3D) paraxial wave equation (PWE), in Cartesian geometry. The source term in the PWE is given by the solution of the Schrödinger equation, in the strong field approximation (SFA)  \cite{LewensteinPRA1994}. The PWE is solved for each spectral component, using a finite-difference method  \cite{CamperPRA2014}, on a 512$\times$512$\times$200 $\mu m^3$ spatial grid and a 100\,fs time interval, for 513$\times$513$\times$41 points in space and 4096 in time. This leads to a spatial step of 1\,$\mu m$ along the transverse dimensions ((x,y) coordinates) and 5\,$\mu m$ in the propagation direction (along the z-axis). The time step is $2.4\times 10^{-2}\,fs$, which is about 1/100th of the optical period of the driving fields. We consider two Gaussian beams of 100\,$\mu m$ waist at focus, i.e. 39\,mm Rayleigh range. The temporal intensity profiles have $\sin^4$ shapes of 50\,fs full-width at half-maximum. Both beams are focused in the middle of a 100\,$\mu m$ thin slab of argon gas, where they spatio-temporally overlap, following the setup depicted in Fig. \ref{Fig:Schematic}. The two beams cross each other with a 2.3$^\circ$ angle. The total peak intensity at focus is $1.5\times 10^{14} W/cm^2$, whatever $\alpha$, and the density of atoms is $3.0\times 10^{17} atoms/cm^{3}$.
\begin{figure}[!htp]
\centering
\includegraphics[width=0.5\textwidth]{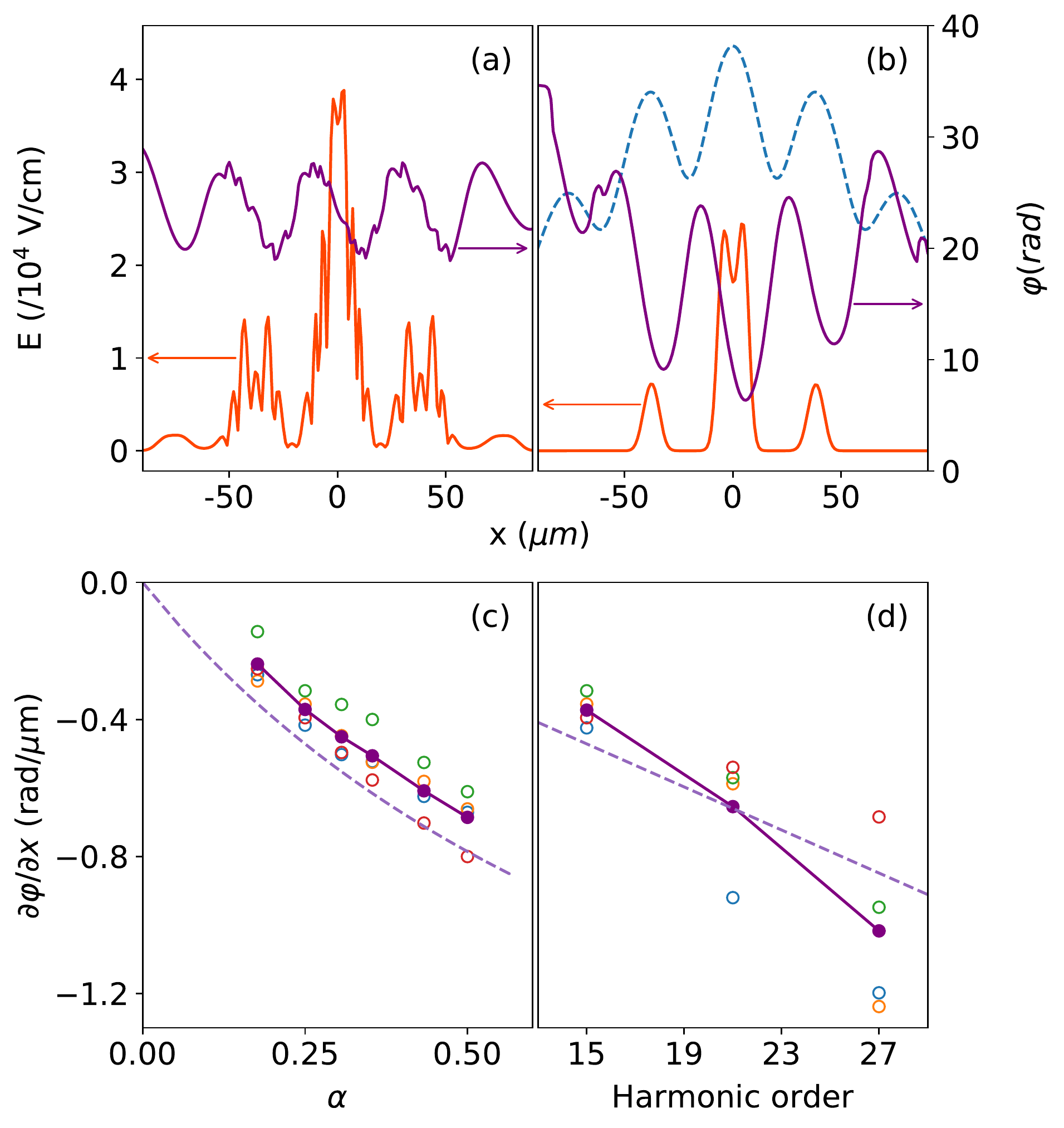}
\caption{(top line) Amplitude (orange) and phase (purple) of harmonic 15 (a) and harmonic 27 (b) at the exit of the generating medium calculated with the model described in Sec. \ref{QuantumModel} versus the transverse direction x. The left (resp. right) scale is used for amplitude (resp. phase) for both panels. In panel (b), the amplitude of the driving field is displayed as a dashed line. It is offset by $2\times 10^8V/cm$ and divided by $10^4$. (bottom line) Values of the derivative of the phase against x. In both plots various colors are results taken for successive maxima of the XUV field (blue: $x\simeq -40\mu m$, orange: $x\simeq 0$, green: $x\simeq 40\mu m$, red: $x\simeq 80\mu m$). The mean value is displayed in purple. (c) Variation of the blaze angle against the field ratio for harmonic q=15. The dashed line is the analytical formula given by Eq. \eqref{eq:phi_0}. (d) Variation of the blaze angle against the harmonic number for $\alpha=0.25$.    }
\label{Fig:FigCutsTA}
\end{figure}
Transverse cuts of the harmonic fields amplitudes and phases are displayed in Fig. \ref{Fig:FigCutsTA} (a-b) for H15, which is in the plateau and H27, which falls in the cutoff for this intensity. As expected, both H15 and H27 show a series of peaks spaced by the period of the active grating (the shape of the driving field is displayed in dashed line). They are located on the highest intensity spots of the driving interference field. It is purely symmetric about $x=0$. In addition, H15 is strongly modulated. This is due to interferences between the long and short trajectories for harmonics lying in the plateau. As a support of our analytical model described above, the phase of the harmonics shows a sawtooth pattern. A linear fit of the phase about the locii of highest intensities gives the corresponding blaze angle, $\partial \varphi/\partial x$. It is displayed against the perturbation ratio $\alpha$ for H15 in Fig. \ref{Fig:FigCutsTA}.c and against the harmonic order for $\alpha=0.25$ in Fig. \ref{Fig:FigCutsTA}. The determination of $\partial \varphi/\partial x$ depends on the ``groove'' considered in the active grating. We displayed the values obtained for four peaks, showing significant dispersion, along with their mean value. Although we did not investigate it further, this dispersion is probably reminiscent of the variations of the $\alpha$ ratios within the focus, as identified in Fig. \ref{Fig:HistoGaussian}. However, the trend in Fig. \ref{Fig:FigCutsTA}.(c) follows reasonably the prediction of our toy model (Eq. \eqref{eq:phi_0} displayed as a dashed line). It should be noted that here, there is no adjustable parameter on the model. The fact that the result is slightly offset upwards is a consequence of the distribution of wave vectors angles above the limit value in Fig. \ref{Fig:HistoGaussian}.(d), especially for harmonics of the plateau that can be generated with fields amplitudes below the peak amplitude. The same conclusions hold for the expected linearity of the harmonic-dependent blaze angle (Fig. \ref{Fig:FigCutsTA}.(d)), further supporting our interpretation of the origin of the offset dominant diffraction orders. These full calculations validate our analytical approach exposed in Sec. \ref{Sec:blazedGrating}. 

\begin{figure}[!htp]
\centering
\includegraphics[width=0.5\textwidth]{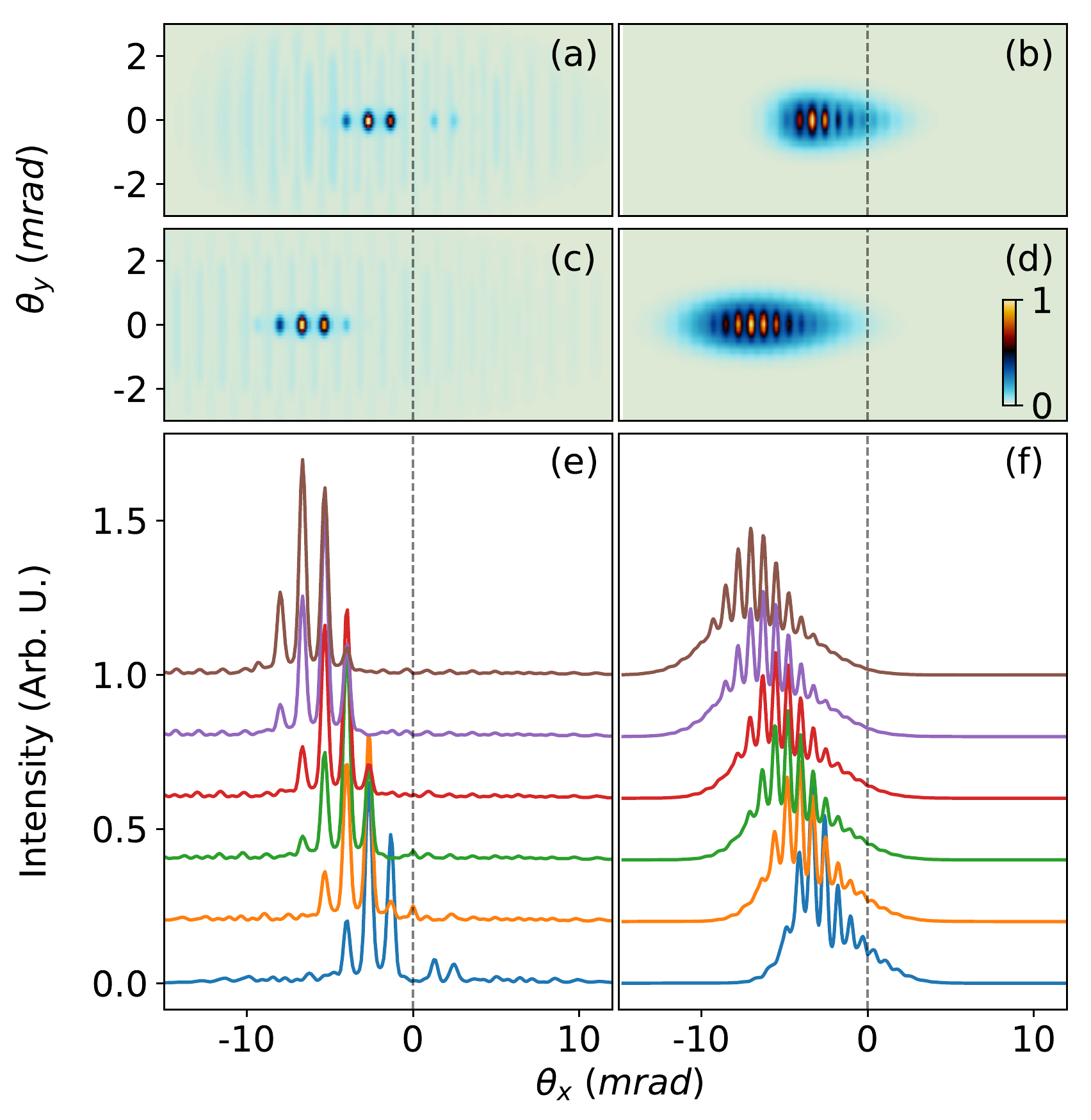}
\caption{Color maps of the intensities in the far field of harmonics 15 (a,c) and 27 (b,d) for perturbations $\alpha=0.18$ (a-b) and $\alpha=0.5$ (c-d). All maps are individually normalized and the color maps are set equal. Cuts of the harmonics intensities 15 (e) and 27 (f) at $\theta_y=0$ for $\alpha=0.18$ (blue), $\alpha=0.25$ (orange), $\alpha=0.31$ (green), $\alpha=0.35$ (red), $\alpha=0.43$ (purple), $\alpha=0.5$ (brown). The successive cuts are offset by 0.2 for the sake of visibility. The intensities of H15 are divided by 2 to share the y-axis with H27.      }
\label{Fig:FigMapsTA}
\end{figure}

To go further, we propagated the harmonic fields towards the far field using the Fresnel propagator. Four intensity maps are displayed in Fig. \ref{Fig:FigMapsTA}, corresponding to the two same harmonics 15 (left column) and 27 (right column), and a weak (top line) and strong perturbation (second line) ($\alpha=0.18$ and $\alpha=0.5$). As anticipated with the analytical model, the harmonics show a series of diffraction orders, more sparse for H15 than H27, below a global envelope. They are all the more shifted towards the SFG side as $\alpha$ is set stronger.  
This is even more evident in Fig. \ref{Fig:FigMapsTA}(e-f) where 6 values of $\alpha$ are used. The similarity of this figure with Fig. \ref{Fig:FigCuts}.(h) is striking. The locations of the dominant peaks even fit fairly well. It should be noted that a slight left/right asymmetry appears, especially for low perturbation values on H27. This is probably a consequence of the extension of the medium along z, which yields such a slight asymmetry of the field. Whereas opposite areas located upstream and downstream the focus are just mirror images about the \emph{x}-axis for the driving field when no gas is inserted, this is no longer the case when reshaping of the fundamental is authorized, nor when harmonics are propagated in differently ionized absorbing media. More careful examination of these ``volume'' effects are left for further studies, the agreement being already extremely promising.

\section{Experimental test of the theory}\label{Sec: exp}
\begin{figure}[htp]
\centering
\includegraphics[width=0.5\textwidth]{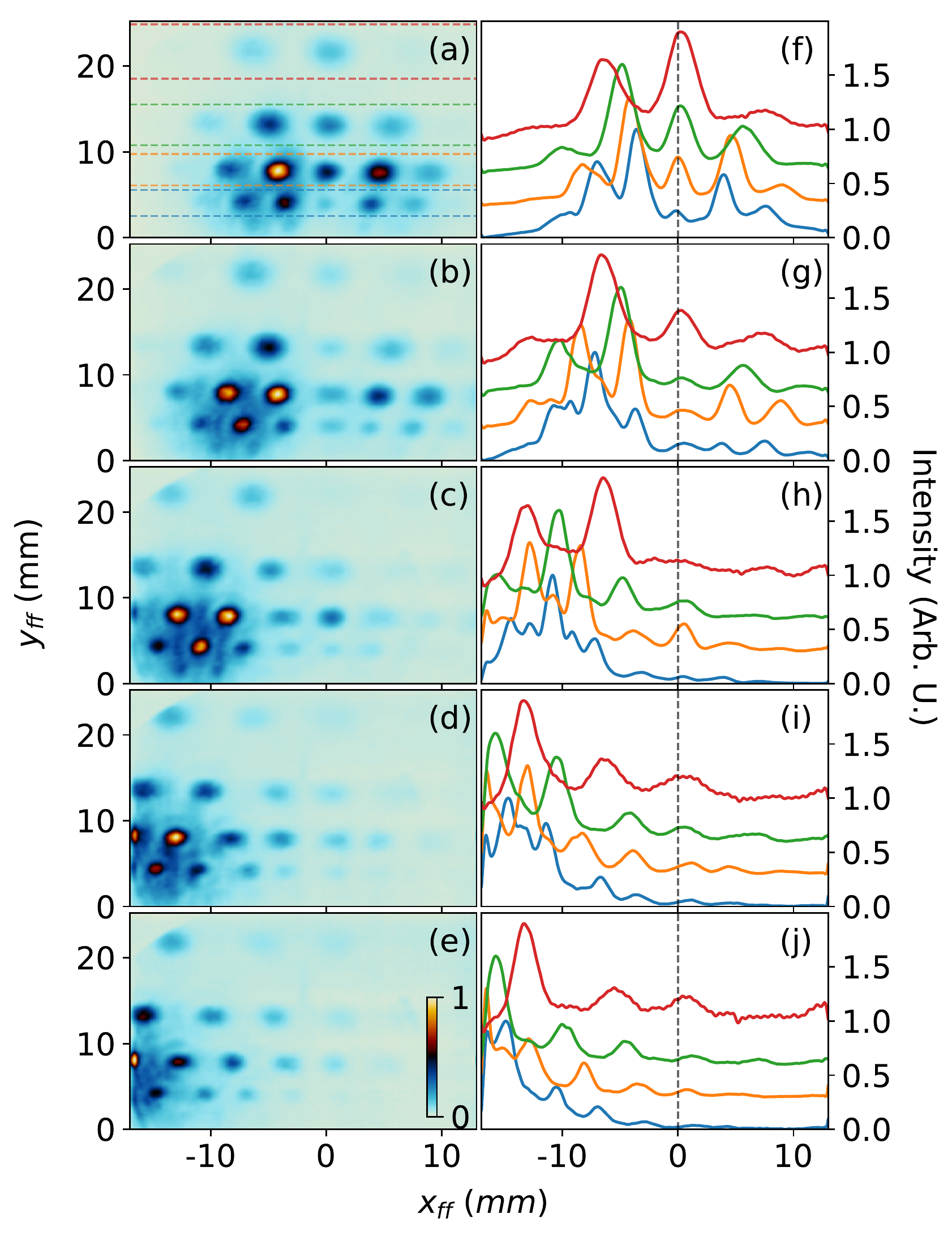}
\caption{(Left column) Experimental images on the detector for five levels of perturbation ($\alpha\lesssim 0.15$, $\alpha= 0.25$, $\alpha=0.38$, $\alpha=0.50$ and $\alpha=0.66$ from (a) to (e)). Here $\theta=1^\circ$. All images are normalized to 1, and share the same colormap. The vertical axis is the direction of the dispersion of the detector grating, while the horizontal axis is the direction of dispersion of the ``active'' grating. (right column) Lineouts corresponding to the sums between the dashed lines displayed in panel (a), for H9 (red), H11 (green), H13 (orange) and H15 (blue). All curves are normalized and offset by 0.3 for the sake of visibility. }
\label{Fig:FigMapsCC}
\end{figure}

We tested the conclusions of these analyses performing experiments on LUCA laser in Saclay. It is a Titanium:sapphire  femtosecond laser based on chirped pulse amplification. It delivers pulses of $\simeq 40$\, mJ energy, 60\,fs full width half maximum (FWHM) duration, at a repetition rate of 20 Hz. It was split into two equal parts which were passed through adjustable attenuators before being focused in an Argon gas jet by two identical lenses of 1\,m focal length. The main beam carried 2.3\,mJ, while the second was adjusted during the experiment to scan $\alpha$. The two beams were linearly polarized vertically and crossed with an adjustable angle in the HHG medium. We focused both beams as close as possible to the gas jet along z. The harmonics generated were collected on a low dispersion grating  \cite{Ruf2012}, before being detected on micro-channel plates (MCP) coupled to a phosphor screen imaged on a CCD camera. The images displayed in Fig. \ref{Fig:FigMapsCC} were averaged over 500 shots. Harmonic numbers were calibrated using the theoretical dispersion of the grating  \cite{geneaux2016}. 
A given harmonic shows at a given y, while the divergence of the harmonics is imaged along the x dimension \footnote{Experimentally, y-axis was horizontal, while x-axis was vertical. }. The left side of the image is cut due to the  size of our MCP set. 

\begin{figure}[!htp]
\centering
\includegraphics[width=0.5\textwidth]{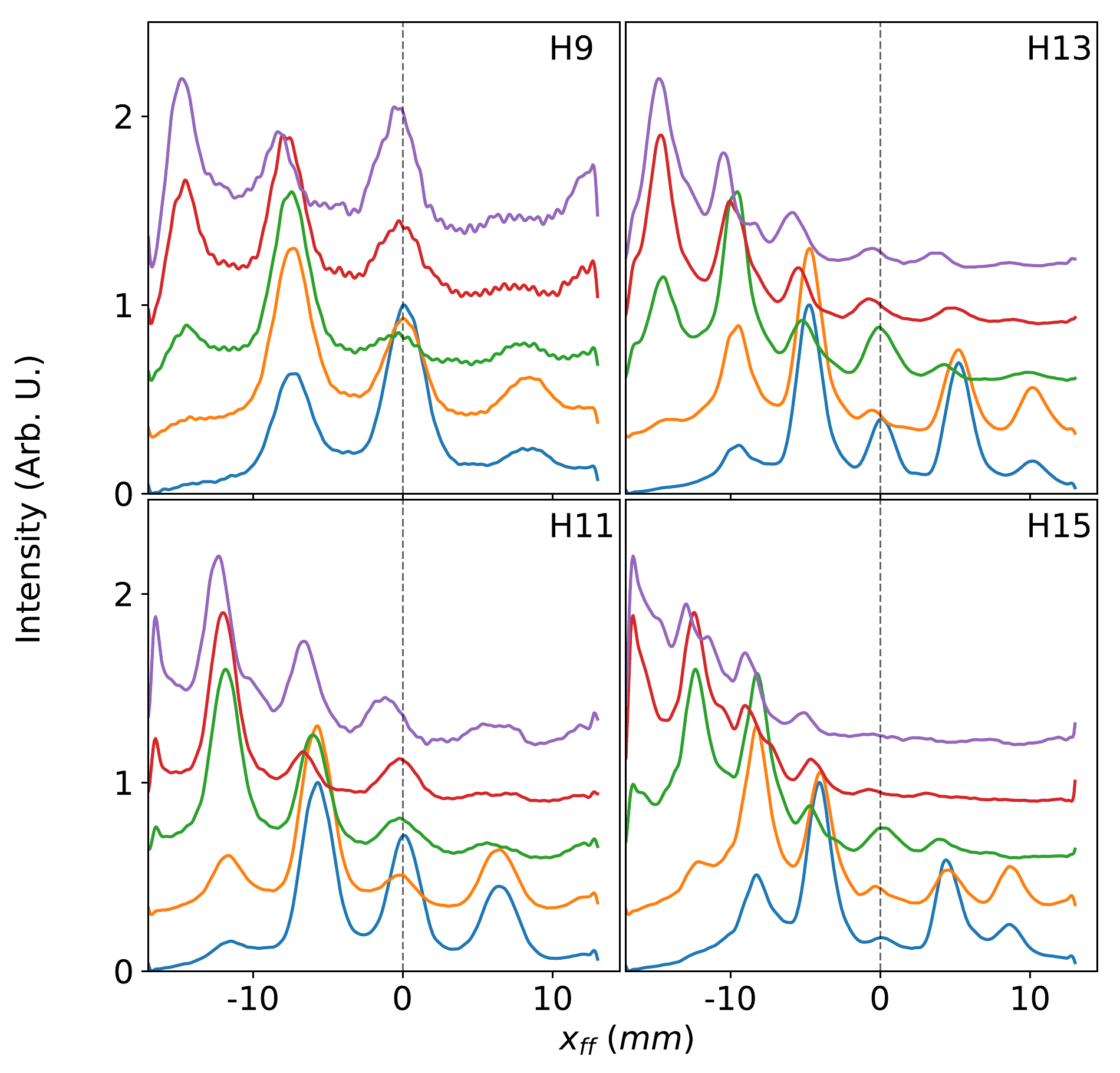}
\caption{Lineouts of experimental images similar to those of Fig. \ref{Fig:FigMapsCC}, but for an angle $\theta=1.2^\circ$. Each panel corresponds to a given harmonic and all curves are normalized to 1 and progressively offset by 0.3. Different colors correspond to perturbation levels of $\alpha\lesssim 0.15$ (blue), $\alpha= 0.25$ (orange), $\alpha=0.38$ (green), $\alpha=0.50$ (red) and $\alpha=0.66$ (mauve)). The dashed vertical line is set at the location of the direct harmonics.    }
\label{Fig:FigCutsCC}
\end{figure}

We clearly observed the spreading of the harmonics into several orders. While both negative and positive orders coexist for low perturbation values, the profiles are quickly deported towards the left of the figure, corresponding to the SFG region. This is all the more the case as $\alpha$ increases (from top to bottom). Interestingly, there are not many more orders appearing as $\alpha$ increases. This is a result predicted by our toy model, where the number of diffraction orders visible is set by the width of the large Gaussian given in Eq. \eqref{eq:widthLargeGaussian}, and confirmed by the full computations. As expected, the progressive shift is compatible with Eq. \eqref{eq:OffsetOrders}. In particular, the lineouts displayed on the right show, for a given panel,  the diffraction orders of all harmonics peaking about the same \emph{x}-location (\emph{e.g.} about -10\,mm for Fig. \ref{Fig:FigCutsCC}(h)). This location is progressively shifting from Fig. \ref{Fig:FigCutsCC} (f) to Fig. \ref{Fig:FigCutsCC} (j) as the slope of the phase grating is increased by increasing the perturbation intensity (increase of $\alpha$ in Eq. \eqref{eq:phi_0}). 
We also retrieve a prediction of Eq. \eqref{eq:LocationOrders}: diffraction orders are getting denser as q increases, which is evident in Fig. \ref{Fig:FigCutsCC}. Finally, we note that a slight left-right asymmetry (e.g. green curve, H13, Fig. \ref{Fig:FigCutsCC}). This is compatible with the full computations but not the toy model. We believe that it is the result of ``volume'' effects. 

\begin{figure}[!htp]
\centering
\includegraphics[width=0.5\textwidth]{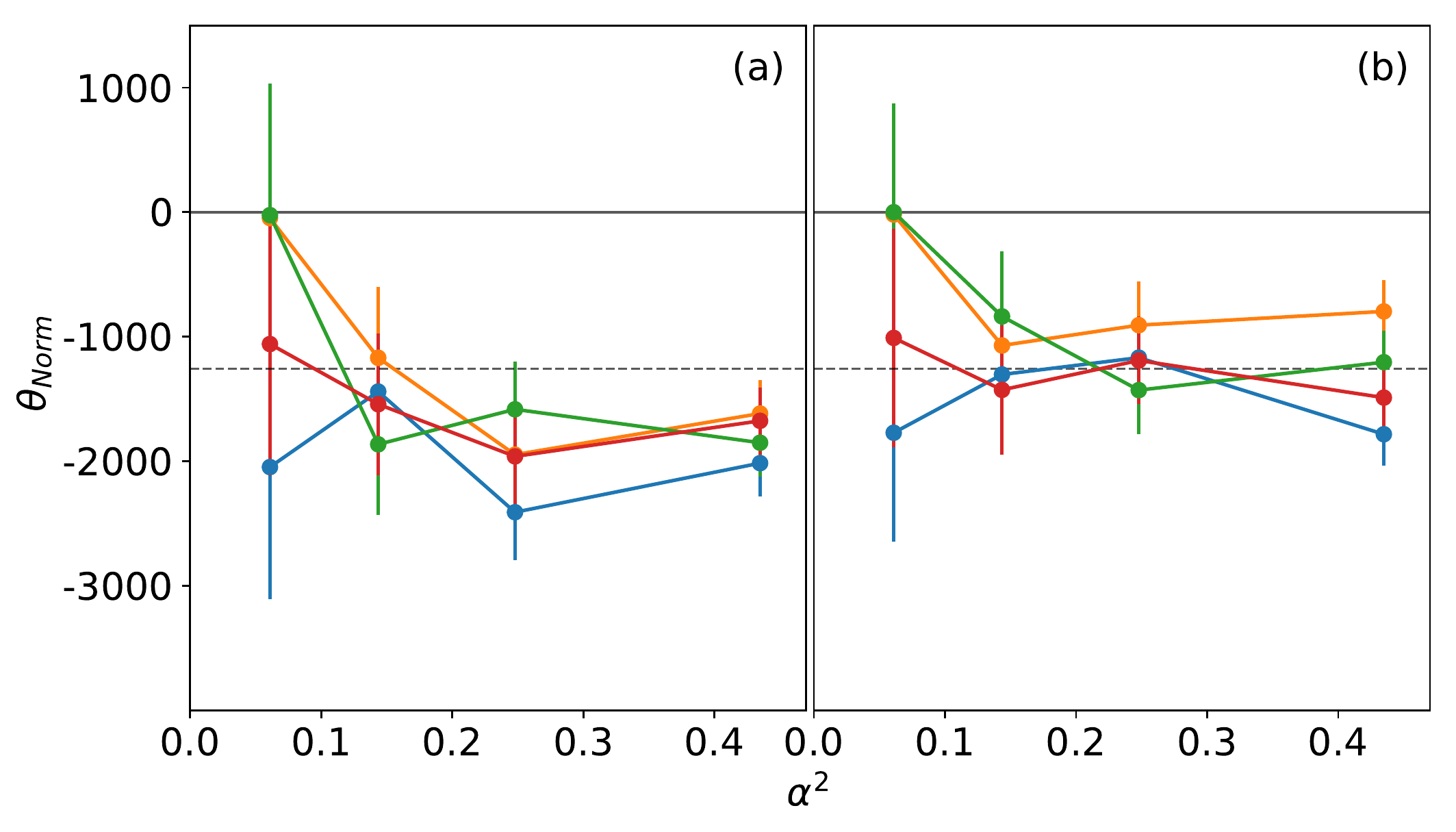}
\caption{Normalized position of the dominant diffraction order as a function of the perturbation energy normalized to the energy in the main beam. Angles $\theta=1^\circ$ (a) and $\theta=1.2^\circ$ (b). Each color corresponds to one harmonic (H9: blue, H11: orange, H13: green, H15: red. The dashed line is the analytical prediction according to Eq. \eqref{eq:ThetaNorm}.   }
\label{Fig:FigNormalized}
\end{figure}

Finally, we tested the formula of the toy model Eq. \eqref{eq:OffsetOrders} which predicts the position of the dominant order. We define 
\begin{equation}
\theta_{Norm}=\frac{x_c'} {z_{ff}}\cdot\frac{1}{\sin\theta}\frac{1+\alpha}{\alpha}.
\label{eq:ThetaNorm}
\end{equation} 
which is the direction of propagation of the harmonics ($\frac{x_c'} {z_{ff}}$) normalized by the perturbation and angle-dependent factor of Eq.  \eqref{eq:OffsetOrders}. It should be a constant whatever the harmonic, angle and perturbation. We plotted it in Fig. \ref{Fig:FigNormalized}, together with the experimental prediction. The error bars here correspond to half the periodicity of the diffraction orders. The agreement is very good considering the simplicity of the model and the fact that no adjustable parameter is available. Except for the first point which suffers from high uncertainty, the normalized angle of propagation is rather constant whatever the harmonic and the perturbation level. It is close to the predicted value. It should be noted that, as a source of uncertainty, we could not perfectly control the overlap in time and space of the two beams for each point. 

\section{Concluding considerations}
With the ever increasing control of high power femtosecond lasers over the years, the availability of multiple beams to perform extreme non-linear optics has been  developing at an extremely rapid pace. Many schemes have been employed, either using a single wavelength or several wavelengths, a single polarization or different polarizations, a gaussian beam or beams carrying orbital angular momenta, in collinear or non collinear geometries, with even counter propagating waves. In the vast majority of these studies, a photonic picture was put forward to interpret the results. As a consequence, ``selection rules'' were derived. They surely reveal which channels may exist. However they do not say anything about possible yields of the concurrent non linear processes allowed. Getting back to the very nature of High Harmonic Generation process, which is a strong field effect and not a multi-photon process, we here exposed an analysis of the process at the ``mesoscopic'' spatial scale, corresponding to several wavelengths. In the specific and simple case of two linearly polarized pulses with identical wavelengths crossing at an angle in the HHG medium, we solved an emerging controversy about the yield of the sum vs different frequency generation processes: SFG processes are dominant over DFG as soon as the ratio between the fields exceeds 10-20\%. We retrieve the \emph{``photonic picture''} as a consequence of the quasi periodicity of the interference pattern in the transverse direction. The excellent agreement of our toy model, full quantum computation and experiment confidently form the basis for future work targeting some of the cases listed above, where the polarizations, wavelengths, angles of crossing, orbital angular momenta may be varied. 


Finally, we point out that the analysis drawn above may have extremely rich applications in the two usual approaches used to probe attosecond dynamics. First, it provides a framework to analyze the amplitudes of high harmonics in high harmonic spectroscopies. It offers an improved understanding of the relative yields covering the perturbative to the non perturbative regimes. Second, it offers the tools to design  the driving field at focus in order to efficiently generate specific harmonics with given properties to be used \emph{ex situ}. In particular, we may envision tailoring the shape of the grating grooves, in phase and amplitude, to generate efficiently a given harmonic, or a set of them. As a more general outlook, a connection to the very thorough and general framework proposed by Bahabad et al. should be built (e.g.  \cite{Bahabad2010, Hareli2018}), together with the incorporation of temporal ``grooves'' as  proposed in the attosecond lighthouse schemes  \cite{KimNP2013,QuereJPB2014,Auguste2016}.This series of outlooks suggest that this work will open new avenues for the investigation of highly non linear processes and the synthesis of smart XUV femtosecond and attosecond pulses.  

\acknowledgments{Financial support from the Agence Nationale pour la Recherche (under contracts ANR11-EQPX0005-ATTOLAB and ANR-14-CE32-0010 - Xstase), the LABEX PALM (ANR-10-LABX-0039-PALM) and the R\'egion Ile-de-France (under contract SESAME-2012-ATTOLITE) is gratefully acknowledged.}
\appendix*
\section{Derivation of the main formula}\label{AppendixFormulas}
\subsection{Derivation of Eq. \eqref{eq:wavevector}}
The local wave vector is defined as the gradient of the phase: 
\begin{equation}
\vec{k}_s(\vec{r})=\vec{\nabla}\left(\vec{k}_1 \vec{r}+\varphi(\alpha,\vec{\Delta k\cdot \vec{r}})\right)
\label{eq:A1}
\end{equation}
with $\varphi(\alpha,\vec{\Delta k\cdot \vec{r}})$ given by Eq. \eqref{AnalForm2}. We thus have
\begin{align}
\begin{split}
\vec{k}_s(\vec{r})-\vec{k}_1={}&\frac{1}{1+\frac{\alpha^2\sin^2(\Delta k\cdot \vec{r})}{\left(1+\alpha\cos(\Delta k\cdot \vec{r})\right)^2}}\cdot  \\
&\left[\frac{\alpha \vec{\Delta k}\cos(\Delta k\cdot \vec{r})}{1+\alpha\cos(\Delta k\cdot \vec{r})}+
\frac{\alpha^2 \vec{\Delta k}\sin^2(\Delta k\cdot \vec{r})}{\left(1+\alpha\cos(\Delta k\cdot \vec{r})\right)^2}\right]
\end{split}\\
\begin{split}
={}&\frac{\alpha \vec{\Delta k} }{\left(1+\alpha\cos(\Delta k\cdot \vec{r})\right)^2+\alpha^2\sin^2(\Delta k\cdot \vec{r})}\cdot \\
&\left[\left(1+\alpha\cos(\Delta k\cdot \vec{r})\right)\cos(\Delta k\cdot \vec{r})+\alpha \sin^2(\Delta k\cdot \vec{r})\right]
\end{split}\\
={}&\frac{\alpha\left(\alpha+\cos(\Delta k\cdot \vec{r})\right)}{1+\alpha^2+2\alpha\cos(\Delta k\cdot \vec{r})}\vec{\Delta k}
\end{align}

\subsection{Derivation of Eq. \eqref{eq:StrongPertub}}
Using $\beta=1-\alpha$ with $\beta\ll 1$ in Eq. \eqref{eq:wavevector}, we get, up to first order in $\beta$
\begin{align}
\vec{k}_s(\vec{r})&=\vec{k}_1+\frac{\alpha \left(\alpha+\cos( \vec{\Delta k}\cdot \vec{r})\right)}{1+\alpha^2+2\alpha\cos( \vec{\Delta k}\cdot \vec{r})}\vec{\Delta k}\\
&= \vec{k}_1+\frac{(1-\beta) \left((1-\beta)+\cos( \vec{\Delta k}\cdot \vec{r})\right)}{1+(1-\beta)^2+2(1-\beta)\cos( \vec{\Delta k}\cdot \vec{r})}\cdot\vec{\Delta k}\\
&\simeq  \vec{k}_1+\frac{(1-\beta) \left(1+\cos( \vec{\Delta k}\cdot \vec{r})\right)-\beta}{1-\beta+(1-\beta)\cos( \vec{\Delta k}\cdot \vec{r})}\cdot\frac{\vec{\Delta k}}{2}\\
&\simeq  \vec{k}_1+\frac{\vec{\Delta k}}{2}-\frac{\beta}{\left(1-\beta\right)\cdot\left(1+\cos( \vec{\Delta k}\cdot \vec{r})\right)}\cdot\frac{\vec{\Delta k}}{2}\\
&\simeq  \vec{k}_1+\frac{\vec{\Delta k}}{2}-\frac{\beta}{1+\cos( \vec{\Delta k}\cdot \vec{r})}\cdot\frac{\vec{\Delta k}}{2}
\label{eq:SP1}
\end{align}

\subsection{Derivation of Eq. \eqref{Modulus_alpha1}}\label{sub:eqModulusAlpha1}
\begin{align}
\left|\vec{k}_s\right|&=\left|\vec{k}_1+\frac{\vec{\Delta k}}{2}\right|\\
&=\sqrt{k_1^2+\vec{k}_1\cdot\vec{\Delta k}+\frac{\Delta k^2}{4}}
\end{align}
with 
\begin{align}
\vec{k}_1\cdot\vec{\Delta k}&=\left(\vec{k}_2-\vec{k}_1\right)\cdot\vec{k}_1\\
&=-{k_1}^2\left(1-\cos \theta\right)\\
&=-2{k_1}^2\cdot \sin^2\frac{\theta}{2}
\end{align}
and, taking into account $|\vec{k_1}|=|\vec{k_2}|$,
\begin{align}
{\vec{\Delta k}}^2&=\left(\vec{k}_2-\vec{k}_1\right)^2\\
&=2{k_1}^2-2{k_1}^2\cos \theta\\
&=4{k_1}^2\cdot \sin^2\frac{\theta}{2}.
\end{align}
We thus get, for small $\theta$
\begin{align}
\left|\vec{k}_s\right|&=k_1\sqrt{1-2\sin^2\frac{\theta}{2}+\sin^2\frac{\theta}{2}}\\
&=k_1\sqrt{1-\sin^2\frac{\theta}{2}}\\
&\simeq  k_1-\frac{k_1 }{2}\sin^2\frac{\theta}{2}.
\end{align}
Finally, 
\begin{align}
\frac{\left|\vec{k}_s\right|-k_1}{k_1}&\simeq - \frac{1 }{2}\sin^2\frac{\theta}{2}.
\end{align}

\subsection{Derivation of Eq. \eqref{Modulus_alpha0}}
Derivations similar to those of Section \ref{sub:eqModulusAlpha1} yield
\begin{align}
\left|\vec{k}_s\right|&=\left|\vec{k}_1+\alpha \vec{\Delta k}\cos(\Delta k\cdot \vec{r})\right|\\
&=\sqrt{k_1^2+2\alpha\vec{k}_1\cdot\vec{\Delta k}\cos(\Delta k\cdot \vec{r})+\alpha^2\Delta k^2 \cos^2(\Delta k\cdot \vec{r})}\\
&\simeq k_1 \left(1-2\alpha\sin^2\frac{\theta}{2}\cos(\Delta k\cdot \vec{r})\right)
\end{align}
Which yield
\begin{align}
\frac{\left|\vec{k}_s\right|-k_1}{k_1}&\simeq - 2\alpha\sin^2\frac{\theta}{2}\cos(\Delta k\cdot \vec{r})\\
&\simeq -\frac{\alpha}{2}\theta^2\cos(\Delta k\cdot \vec{r})
\end{align}

\subsection{Derivation of Eq. \eqref{Angle_Alpha0}}
From Eq. \eqref{eq:LocalLowAlpha}, the angle of the wave vector associated to the sum of the two fields is 
\begin{equation}
\theta_{s}=\arctan{\frac{\alpha k_2\sin \theta}{k_1-\alpha\cdot(k_1-k_2\cos\theta)\cos\left(\vec{\Delta k}\cdot \vec{r}\right)}}.
\label{eq:thetasumA1}
\end{equation}
  Taking into account that $k_2=k_1$ and $\alpha \ll 1$, we get
\begin{align}
\theta_{s}&=\arctan{\frac{\alpha \sin \theta\cos\left(\vec{\Delta k}\cdot \vec{r}\right)}{1-\alpha\cdot(1-\cos\theta)\cos\left(\vec{\Delta k}\cdot \vec{r}\right)}}\label{eq:thetasumA2}\\
&\simeq \alpha \sin \theta\cos\left(\vec{\Delta k}\cdot \vec{r}\right)\label{eq:thetasumA3}
\end{align}

\subsection{Derivation of Eq. \eqref{Modulus_alphaGeneral}}
\begin{align}
\left|\vec{k}_s\right|&=\left|\vec{k}_1+\frac{\alpha}{1+\alpha}\vec{\Delta k}\right|\\
&=\sqrt{k_1^2+2\frac{\alpha}{1+\alpha}\vec{k}_1\cdot\vec{\Delta k}+\frac{\alpha^2}{\left(1+\alpha\right)^2}\Delta k^2}
\end{align}
Using the expression derived in \ref{sub:eqModulusAlpha1} for $\vec{k_1}\cdot\vec{\Delta k}$ and ${\Delta k}^2$ we get 
\begin{align}
\left|\vec{k}_s\right|&=k_1\sqrt{1-4\frac{\alpha}{1+\alpha}\sin^2\left(\frac{\theta}{2}\right)+\frac{\alpha^2}{\left(1+\alpha\right)^2}\cdot4\sin^2\left(\frac{\theta}{2}\right)}\\
&=k_1\sqrt{1-\frac{4\;\alpha}{\left(1+\alpha\right)^2}\cdot\sin^2\left(\frac{\theta}{2}\right)}
\end{align}

\subsection{Derivation of Eq. \eqref{Angle_General}}
The angle of the wave vector associated to the sum of the two fields is 
\begin{equation}
\theta_{s}=\arctan{\frac{\frac{\alpha}{1+\alpha} k_2\sin \theta}{k_1-\frac{\alpha}{1+\alpha}\cdot(k_1-k_2\cos\theta)}}.
\label{eq:thetasumGeneral1}
\end{equation}
Taking into account $|\vec{k_1}|=|\vec{k_2}|$,
\begin{align}
\theta_{s}&=\arctan{\left(\frac{\alpha}{1+\alpha}\cdot\frac{\sin \theta}{1-\frac{\alpha}{1+\alpha}\cdot(1-\cos\theta)}\right)}\label{eq:thetasumA4}\\
&=\arctan{\left(\frac{\sin \theta}{\frac{1+\alpha}{\alpha}-(1-\cos\theta)}\right)}\label{eq:thetasumA5}\\
&=\arctan{\left(\frac{\alpha\sin \theta}{1+\alpha\cos\theta}\right)}\label{eq:thetasumA6}
\end{align}



\bibliographystyle{unsrt}
\bibliography{C:/AAA_Donnees/Library/Reflib}

\end{document}